\documentclass[british,english, nofootinbib, aps]{revtex4}
\usepackage[T1]{fontenc}
\usepackage[latin9]{inputenc}
\usepackage{xcolor}
\usepackage{pdfcolmk}
\usepackage{amsmath}
\usepackage{graphicx}
\usepackage{esint}
\PassOptionsToPackage{normalem}{ulem}
\usepackage{ulem}
\makeatletter
\providecolor{lyxadded}{rgb}{0,0,1}
\providecolor{lyxdeleted}{rgb}{1,0,0}

\@ifundefined{textcolor}{}
{%
 \definecolor{BLACK}{gray}{0}
 \definecolor{WHITE}{gray}{1}
 \definecolor{RED}{rgb}{1,0,0}
 \definecolor{GREEN}{rgb}{0,1,0}
 \definecolor{BLUE}{rgb}{0,0,1}
 \definecolor{CYAN}{cmyk}{1,0,0,0}
 \definecolor{MAGENTA}{cmyk}{0,1,0,0}
 \definecolor{YELLOW}{cmyk}{0,0,1,0}
}

\usepackage{color, epsfig, graphics}
\usepackage{epstopdf}
\@ifundefined{showcaptionsetup}{}{%
 \PassOptionsToPackage{caption=false}{subfig}}
\usepackage{subfig}
\makeatother

\usepackage{babel}
\begin{document}

\title{Arbitrary axisymmetric steady streaming: Flow, force and propulsion}

\author{Tamsin A. Spelman}

\author{Eric Lauga}

\email{e.lauga@damtp.cam.ac.uk}

\address{Department of Applied Mathematics and Theoretical Physics, University
of Cambridge, Wilberforce Road, Cambridge, CB3 0WA, United Kingdom}

\date{\today }

\begin{abstract}
A well-developed method to induce mixing on microscopic scales is to exploit flows generated by steady streaming. Steady streaming is a classical fluid dynamics phenomenon whereby
a time-periodic forcing in the bulk or along a boundary is enhanced
by inertia to induce a non-zero net flow. Building on classical work
for simple geometrical forcing and motivated by the complex shape
oscillations of elastic capsules and bubbles, we develop the mathematical
framework to quantify the steady streaming of a spherical body with
arbitrary axisymmetric time-periodic boundary conditions. We compute
the flow asymptotically for small-amplitude oscillations of the boundary
in the limit where the viscous penetration length scale is much smaller
than the body. In that case, the flow has a boundary layer structure
and the fluid motion is solved by asymptotic matching. Our results,
presented in the case of no-slip boundary conditions and extended
to include the motion of vibrating free surfaces, recovers classical
work as particular cases. We illustrate the flow structure given by
our solution and propose one application of our results for small-scale
force-generation and synthetic locomotion. 
\end{abstract}
\maketitle

\section{Introduction}

Flow mixing and transport on small scales have been widely studied  \cite{squires05}. These processes can be non-trivial as small-scale flows often occur at low Reynolds number and as such are laminar and  time-reversible.  Achieving  precise flow and mixing control is however critical experimentally in a range of
medical and chemical applications \cite{SiaWhitesides_MicroDevicesBiology2003}.
Applying pressure differences or inducing electrokinetic flows are the two most common methods
used to drive flows on small scales \cite{squires05} and mixing has often been achieved by encouraging chaotic particle paths e.g.~by designing a device with complex geometries \cite{MicrofluidicsApplilcations_Stone,SiaWhitesides_MicroDevicesBiology2003}. 

As an alternative, one can look at the biological world for inspiration in design since mixing and transport occurs  on cellular length scales \cite{MesoScaleTurbulence_Ray}.  
Individual microorganisms are known to generate a net dipolar flow in the surrounding fluid when they self-propel  \cite{lauga2009,MeasurementOfFlowField_Ray}. A population of swimmers generates larger more varied flows, which can aid mixing, through a variety of processes including bioconvection \cite{HydrodynamicOrganismsReview_Tim} and purely  collective effects \cite{saintillan08,BacterialTurbulence_Ray}.  More subtly their presence may affect the diffusive transport of passive tracers which develops non-Gaussian tails in the presence of biological activity \cite{EnhancedTrancer_Kyr&Ray,StirringSquirmers_Thiffeault,Zaid_DiluteSuspensions_2011}.

From a modelling standpoint,  analytic studies of  flows induced by swimmers  is older than experimental observations due to previous limitations of imaging techniques.  Many approximately spherical microorganisms (e.g.~the multicellular alga {\it Volvox} or the protozoon {\it Opalina}) move by waving slender appendages  called cilia. A standard modelling approach consists in  considering the dynamics of the enclosing envelope of the cilia, thus reducing the problem to that of a spherical body inducing a surface wave of deformation \cite{Blake1971,Brennen_CiliaryPropulsion_1974}. The first of such analytic models was proposed by Lighthill \cite{Lighthill1951_Squirmer} 
and later corrected by Blake \cite{Blake1971}, who calculated the net flow generated by small-amplitude axisymmetric oscillations of a spherical surface in a Stokes flow. This model is now refereed to as the ``squirmer'' model.  More recent studies
have extended this model to include non-axisymmetric motion \cite{Lauga2014_GeneralisedSqurimingMotion}, the presence of nearby boundaries \cite{Ardekani2014_SquirmerWallNumerics,Lauga2012-SquirmerWall} 
and large-amplitude oscillations \cite{Ishimoto2014_EffciencyBeyondLighthill}.

Recently the importance of inertia to the swimming of microorganisms
was recognised. Unsteady inertial effects are particularly influential over short time scales \cite{Wange2012_UnsteadySwimmer} but not on longer timescales \cite{Ishimoto2013_UnsteadySquirmer}.
A theoretical extension of  Lighthill's original model to include unsteady inertia
was carried out by Rao \cite{Rao1988_UnsteadyStokes}. Convective
inertia, in contrast,  is important for larger microorganisms,  typically of radius close to one
millimeter. Using a squirmer model with tangential forcing on the fluid only, two recent studies 
considered how convective inertia impacted the swimming kinematics and efficiency, emphasizing in particular the relationship between the direction of the far-field flow and the inertial effects \cite{Wang2012_InertialSquirmer,khair14}

In the lab, a variety of synthetic drivers and mixers at the micron-scale have been
designed and experimentally tested  \cite{Hessel2005_MicromixingReview,Nguyen2005_MixerReview}. The  simplest devices are time-periodic and exploit the
inertia in the fluid to generate net motion. This motion is the classical
phenomenon of steady (or acoustic) streaming, whereby
a time-periodic forcing is nonlinearly rectified
by inertia to induce a non-zero net flow \cite{Riley2001_StreamingReview}. 
One method  to obtain this time periodic forcing is to use microbubbles
whose surfaces oscillate when forced with ultrasound, allowing the transformation of 
a macroscopic piezoelectric forcing into  oscillations at the  micrometer scale,  with a range of potentially important biomedical applications  \cite{BubbleBioAppRev_2014}. 
Studies of the streaming generated by oscillating microbubbles started with  Elder's experiments in 1958 \cite{Elder1958_Microstreaming}. Microbubbles
are able to  transport particles either in their streaming flow  when
the microbubble is held stationary  \cite{Marmottant_2004_MicroTransport,Marmottant2003_Nature}
or where the microbubble itself carries the particle \cite{BubbleBioAppRev_2014,Fend_Micropropulsion_2015}. Interactions between multiple oscillating bubbles may be used to increase mixing flows  \cite{Wang2013_FrequencyControl,Ahmed_2009_Mixer,SSWang2009_FreeBubbleMixer}.

Theoretical studies of the steady streaming flows have focused on  shape oscillations in simple geometries, including 
a translating sphere \cite{Riley1966_SphereOscillating}, a translating
bubble \cite{Riley1971_TranslatingBubble,Higgins1998_Bubble}, 
a bubble both translating and pulsating \cite{Higgins1998_Bubble}, 
and more recently a bubble both pulsating and oscillating with one
higher-order Legendre mode \cite{Marksimov2007_GeneralBubble}.  For free microbubbles, the external acoustic energy is focused into the first few surface modes of oscillation hence these classical studies are sufficient  to model streaming. However as  setups become more complicated, for example in the case of  solid capsules  partially enclosing  three-dimensional  bubbles \cite{Fend_Micropropulsion_2015,Ahmed_2015_Micropropulsion,BertinSwimmer_2015}, it is important to be able to model the complex shape dynamics and
accurately compute the resulting streaming flows and forces.

In this paper, we develop the mathematical framework to quantify the steady streaming of a spherical body with arbitrary axisymmetric time-periodic boundary conditions.  We compute the flow asymptotically under two assumption: (1) the amplitude of surface oscillations are small relative to the size of the body (ratio of amplitude $\epsilon \ll 1$); and (2) the acoustic frequency is large such that the viscous penetration length scale is small compared to the body size (ratio penetration length to body size  $\delta\ll1$). Mathematically, for a fixed value of $\delta$, we solve the flow in powers of $\epsilon$, and thus focus implicitly on the limit $\epsilon\ll\delta\ll1$. This limit is the relevant one for micron-size   bubbles    forced by  ultrasound  (frequencies in the hundreds of kHz range) and milimeter-sized organisms. Similarly to classical work, the flow is shown to have a boundary layer structure and the problem is  solved by asymptotic matching.  Our results, presented in the case of no-slip boundary conditions and extended to include the motion of vibrating free surfaces, recovers classical work as particular cases. We then illustrate the flow structure given by our solution and propose  one application of our results on small-scale force-generation and synthetic locomotion. 

%%%%

Our paper is organised as follows. In \S \ref{part:GeneralAcousticStreaming} we set up the problem of the fluid flow generated by arbitrary surface motion of a no-slip spherical body. In \S \ref{sec:First_order_Asymptotic_solution}, we derive the first-order solution. The second-order Eulerian steady  streaming is derived in \S \ref{sec:Second_order_Asymptotic_solution} which is extended to give the Lagrangian steady streaming in \S \ref{sec:LagrangianStreaming}. The special case of a squirming microorganism is then discussed in \S \ref{part:SPECIALCASE_Squirmer}.  This no-slip general model can be extended to incorporate other surface motion such as a no tangential stress boundary as shown in   \S \ref{sec:AllowingBoundarySlip}. 
 The special case of a bubble is then considered in  \S \ref{part:SPECIALCASE_bubble}.   In \S \ref{part:ComparisonRileyHiggins} our general solution is validated against classical results  for  spheres and bubbles.  In \S \ref{part:StreamingExamples}, we illustrate  examples of streaming flows. In all  previous sections, the body was assumed to be held stationary at the origin.  In \S \ref{part:ForceAndVelocity} the time-averaged force induced by the flow on the fixed body is calculated, along with the translational velocity of the spherical body if it was instead free to move.

\section{Axisymmetric steady streaming: Setup}
 \label{part:GeneralAcousticStreaming}
 
In this first section we present the general setup for our calculation. The body is taken to be spherical  with an imposed axisymmetric, radial and tangential time-periodic deformation of its surface. In the following sections we will  first  characterize the flow in the case of no-slip between the fluid and the surface, and  then we generalise to allow the formulation to be  adapted to  other boundary conditions, in particular  no tangential stress for  a clean bubble.

\subsection{Statement of the mathematical problem}

\begin{figure}
\includegraphics[width=0.5\textwidth]{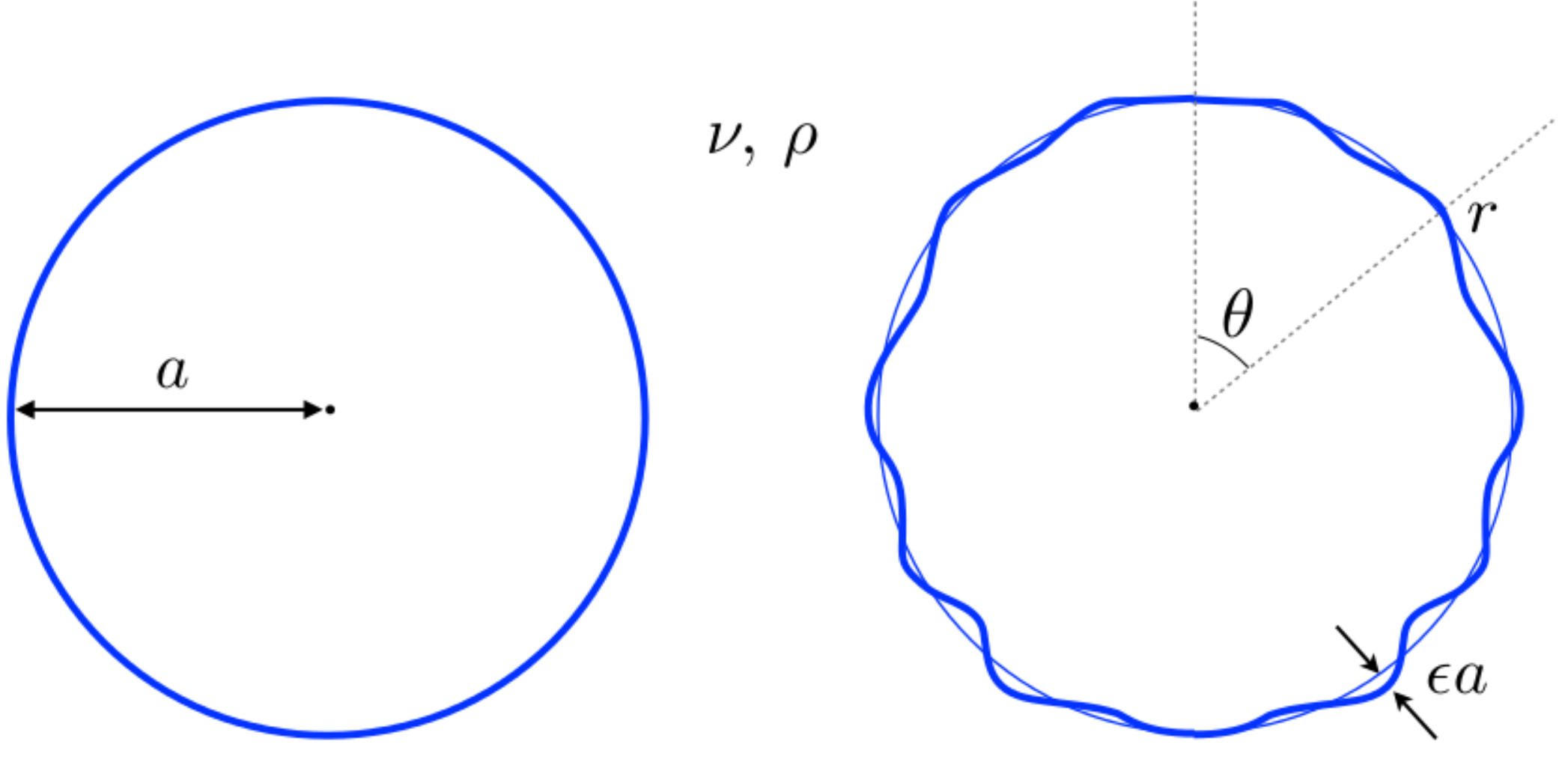}
\caption{A sphere of rest radius $a$ (left) undergoes arbitrary axisymmetric vibrations of amplitude $
\epsilon a$, with $\epsilon\ll 1$ (right). Spherical coordinates are used with  radial distance $r$ and polar angle $ 
\theta$. The surrounding fluid is Newtonian with kinematic viscosity $\nu$ and density $\rho$.}
\label{fig1}
\end{figure}

The sphere has  mean radius $a$  and is contained within
an unbounded Newtonian fluid of constant kinematic viscosity $\nu$ and uniform density 
$\rho$ (Figure~\ref{fig1}). Working in a spherical
coordinate system centred on the sphere, with radial distance $r$
and  polar  angle $\theta$, the axis $\theta=0$ is the axis of rotational
symmetry. The surface of the body is assumed to  oscillate at angular frequency $\omega$ with small amplitude $\epsilon a$, where $\epsilon \ll 1$ is formally specified below. 
 Define $\mu\equiv \cos \theta$, then since the flow is axisymmetric a 
 streamfunction $\psi$ can be introduced
to give radial $u_{r}$ and angular $u_{\theta}$ velocities as
\begin{align}
u_{r} & =-\frac{1}{r^{2}}\frac{\partial\psi}{\partial\mu},\\
u_{\theta} & =\frac{-1}{r(1-\mu^{2})^{\frac{1}{2}}}\frac{\partial\psi}{\partial r}\cdot
\end{align}
The governing equation is then given by the vorticity equation \cite{batchelor1967}
\begin{equation}
\frac{\partial(D^{2}\psi)}{\partial t}+\frac{1}{r^{2}}\left[\frac{\partial(\psi,D^{2}\psi)}{\partial(r,\mu)}+2D^{2}\psi L\psi\right]=\nu D^{2}\left(D^{2}\psi\right),\label{eq:GoverningEquationDimensional}
\end{equation}
where we have defined the operators
\begin{align}
D^{2} & \equiv \frac{\partial^{2}}{\partial r^{2}}+\frac{(1-\mu^{2})}{r^{2}}\frac{\partial^{2}}{\partial\mu^{2}},\label{eq:D2explanation}\\
L & \equiv \frac{\mu}{(1-\mu^{2})}\frac{\partial}{\partial r}+\frac{1}{r}\frac{\partial}{\partial\mu}\cdot
\end{align}

In order to non-dimensionalise the problem, we  take the relevant time scale to be  $\omega^{-1}$ and the relevant length scale to be  $a$, so that the sphere now has a rest radius of 1. The streamfunction thus has dimensions $a^{3}\omega$  and the  vorticity equation becomes
\begin{equation}\label{6}
\frac{\partial(D^{2}\psi)}{\partial t}+\frac{1}{r^{2}}\left[\frac{\partial(\psi,D^{2}\psi)}{\partial(r,\mu)}+2D^{2}\psi L\psi\right]=\left(\frac{\nu}{\omega a^{2}}\right)D^{2}\left(D^{2}\psi\right).
\end{equation}
Equation \eqref{6} introduces a non-dimensional quantity: the ratio of the viscous penetration length scale, $\sim (\nu/\omega)^{1/2}$, to the radius of the body, $a$. Specifically, we define 
a dimensionless number $\delta$ as \begin{equation}
\delta\equiv \frac{\left({2\nu}/{\omega}\right)^{\frac{1}{2}}}{a},\label{eq:delta_boundarylayersize}
\end{equation}
thus reducing the governing equation to 
\begin{equation}
\frac{\partial(D^{2}\psi)}{\partial t}+\frac{1}{r^{2}}\left[\frac{\partial(\psi,D^{2}\psi)}{\partial(r,\mu)}+2D^{2}\psi L\psi\right]=\left(\frac{\delta^{2}}{2}\right)D^{2}\left(D^{2}\psi\right).\label{8}
\end{equation}

The second dimensionless quantity in this problem is the ratio $\epsilon$ between the amplitude of oscillation and the body radius. To use notations similar to those in classical steady streaming calculations, we use  $U$  to denote the maximum velocity at the surface of the oscillating body, such that  $\epsilon$ can be defined as
\begin{equation}
\epsilon=\frac{U}{a\omega}\cdot
\end{equation}

We will look to solve this problem as a regular expansion in  $\epsilon$ for small values of $\delta$ , and then take a regular expansion in $\delta$. We will thus assume the asymptotic limit $\epsilon\ll\delta\ll1$.  Physically, a small value of $\epsilon$ indicates small-amplitude motion while  a small value of $\delta$ means that the viscous penetration length is small compared to the rest size of the body. For which practical situations will these limits be relevant? To fix ideas, let  us take a value for the relative amplitude of $\epsilon\sim10^{-2}$. A ciliated microorganism in water ($\nu=10^{-6}$~m$^2$/s) would have intrinsic frequencies  of about 50 Hz, so $\omega\approx 300$~rad/s, leading to a penetration length of $({2\nu}/{\omega})^\frac{1}{2}\approx 80$~$\mu$m. In order to satisfy the limit  $\epsilon\ll\delta\ll1$, the organism size would need to be just below 1~mm, which is achieved for the largest ciliated organisms such as {\it Spirostomum} which can grow up to 4~mm in length \cite{ishida1988cell}. 
For a micro bubble actuated by ultrasound the frequency is about  $\omega\sim10^{6}$~rad/s, so the penetration length is $({2\nu}/{\omega})^\frac{1}{2}\approx 1$~$\mu$m and thus the bubble would have to be about 10~$\mu$m in diameter.

%%%%
\subsection{Boundary conditions}

We apply, in this first part of the paper, the no-slip boundary condition. Thus, the fluid velocity has to match the velocity of the material points on the surface of the body, which is arbitrary and decomposed along an infinite sum of surface modes.  
Using a Lagrangian formulation, the motion of the boundary can be
described by its radial position, $R$, and angular position measured from the
axis of axisymmetry, $\Theta$, which are  functions of time $t$ and the rest angular position  $\theta$ (through $\mu$) as
\begin{eqnarray}
R&=&1-\epsilon\sum_{n=0}^{\infty}V_{n}P_{n}(\mu)e^{i(t+\frac{\pi}{2})}+O(\epsilon^{2}),\label{eq:RadialPosition}\\
\Theta&=&\theta+\epsilon\sum_{n=1}^{\infty}W_{n}\left(\frac{\int_{\mu}^{1}P_{n}(x)dx}{(1-\mu^{2})^{\frac{1}{2}}}\right)e^{i(t+\frac{\pi}{2})}+O(\epsilon^{2}),\label{eq:AngularPosition}
\end{eqnarray}
where $V_{n}\text{ and }W_{n}$ are arbitrary complex constants determined by
the surface motion of the spherical body  and $P_{n}(x)$ is the
Legendre Polynomial of degree $n$. Throughout the paper,  complex notation will be used and it will always be implied  that only the real part is taken; when an explicit real part appears we will denote it $\Re$.

The $\mu$-dependence was chosen in order to match the form
of the first-order solution, as seen below. Through the many possible choices for constants $V_{n}$ and $W_{n}$, a wide range of boundary motions can be studied. At leading order, $R(\theta)$ is equivalent to the radial position of the surface at an angle $\theta$ from the axis of symmetry. As such, $V_{n}$ will   be determined by the shape of the surface oscillation. At leading order, $\Theta(\theta)$ captures  the tangential motion at an angle $\theta$ from the axis of symmetry, and thus $W_{n}$ will be determined by the appropriate in-surface motion.  We note that the use of   Legendre
polynomials as a basis for $\mu$ was expected due to the problem's
axisymmetry and such a basis has appeared in other work based in similar
regimes \cite{Blake1971}. As the Legendre polynomials form a complete
orthogonal basis, they allow for an arbitrary  solution
to be written in this form.

At $O(\epsilon^{2})$ only terms in $R$ and $\Theta$ which time-average to a non-zero value would contribute to the streaming. However such terms would indicate that the body was slowly growing or shrinking over time, and would also stretch or contract tangentially, violating the small-amplitude assumption on long time scales. We therefore do not allow for steady Lagrangian terms at order $O(\epsilon^{2})$. The boundary condition can thus be written in a Lagrangian form as 
\begin{eqnarray}
u_{r} & =&\frac{\partial R}{\partial t}=\epsilon\sum_{n=0}^{\infty}V_{n}P_{n}(\mu)e^{i t}+O(\epsilon^{3}),\label{eq:u_r}\\
u_{\theta}&=&R\frac{\partial\Theta}{\partial t}=\Re\left[-\epsilon\sum_{n=1}^{\infty}W_{n}\left(\frac{\int_{\mu}^{1}P_{n}(x)dx}{(1-\mu^{2})^{\frac{1}{2}}}\right)e^{it}\right]\Re\left[1-\frac{\epsilon i}{\omega}\sum_{n=0}^{\infty}V_{n}P_{n}(\mu)e^{it}\right]+O(\epsilon^{3}),\label{eq:u_theta}
\end{eqnarray}
and both of which have to be evaluated at  $(r,\theta)=(R(\mu,t),\Theta(\mu,t))$. Finally, we require that the flow decay to zero from the body and thus 
 $u_{r,\theta}\to 0$ as $r\to\infty$.

\subsection{Rearranging the surface boundary conditions}

The current Lagrangian form of the body's surface conditions 
(\ref{eq:u_r})-(\ref{eq:u_theta})  needs to be transformed into Eulerian boundary conditions of the fluid motion.  This is achieved by Taylor expanding them   about the average oscillation position $(1,\theta)$ which in
spherical coordinates gives 
\begin{eqnarray}
\left(\begin{array}{c}
u_{r}\\
u_{\theta}
\end{array}\right)_{r=R}=\left(\begin{array}{c}
u_{r}\\
u_{\theta}
\end{array}\right)_{r=1}&+&\Re\left[(R-1)\right]\Re\left[\left(\begin{array}{c}
\frac{\partial u_{r}}{\partial r}\\
\frac{\partial u_{\theta}}{\partial r}
\end{array}\right)_{r=1}\right]\\
&+&\Re[(\Theta-\theta)]\Re\left[\left(\begin{array}{c}
\frac{\partial u_{r}}{\partial\theta_{0}}-u_{\theta}\\
\frac{\partial u_{\theta}}{\partial\theta_{0}}+u_{r}
\end{array}\right)_{r=1}\right]+O(\epsilon^{2}).\label{eq:TaylorExpansin}
\end{eqnarray}

Looking for the dimensionless velocities and streamfunction as power
series in $\epsilon$ as 
\begin{eqnarray}
u_{r}&=&\epsilon u_{r}^{(1)}+\epsilon^{2}u_{r}^{(2)}+O(\epsilon^{3}),\label{16}\\
u_{\theta}&=&\epsilon u_{\theta}^{(1)}+\epsilon^{2}u_{\theta}^{(2)}+O(\epsilon^{3}),\label{17}\\
\psi&=&\epsilon\psi_{1}+\epsilon^{2}\psi_{2}+O(\epsilon^{2}).
\end{eqnarray}
then at leading order (\ref{eq:TaylorExpansin})  gives 
\begin{eqnarray}
u_{r}^{(1)}&= & \sum_{n=0}^{\infty}V_{n}P_{n}(\mu)e^{it},\label{eq:FirstOrderBoundaryCondition_One}\\
u_{\theta}^{(1)} & = & -\sum_{n=1}^{\infty}W_{n}\left(\frac{\int_{\mu}^{1}P_{n}(x)dx}{(1-\mu^{2})^{\frac{1}{2}}}\right)e^{it},\label{eq:SecondOrderBoundaryCondition_Two}
\end{eqnarray}
at $r=1$ and for all values of $\theta$. 
Simiarly from (\ref{eq:TaylorExpansin}) the $O(\epsilon^{2})$ boundary
condition can be calculated. However due to the non-linear terms arising from  the Taylor expansions,  the first-order flow needs to be  first evaluated in order to determine the
$O(\epsilon^{2})$ boundary conditions explicitly. This will  be discussed in \S\ref{sec:Second_Order_Boundary_Conditions}.

\section{First-order Asymptotic solution\label{sec:First_order_Asymptotic_solution}}

\subsection{General solution}
Based on the oscillatory nature of the boundary condition at first order, \eqref{eq:FirstOrderBoundaryCondition_One}-\eqref{eq:SecondOrderBoundaryCondition_Two}, we look for a solution 
 $\psi_{1}\propto e^{it}$. At leading order the governing equation reduces to 
\begin{equation}
\left(\frac{\partial}{\partial t}-\frac{\delta^{2}}{2}D^{2}\right)(D^{2}\psi_{1})=0.\label{eq:firstordergoverning}
\end{equation}
This is easily solved using separation of variables $D^{2}\psi_{1}=f(r)g(\mu)e^{it}$
to find $D^{2}\psi_{1}$ as 
\begin{equation}
D^{2}\psi_{1}=e^{it}\left[{\scriptstyle {\displaystyle \sum_{n=1}^{\infty}\left(\int_{\mu}^{1}P_{n}(x)dx\right)\left(B_{n}\sqrt{r}K_{n+\frac{1}{2}}(\alpha r)\right)}}\right].\label{eq:D^2*psi_1 Solution}
\end{equation}
where $K_{a}(x)$ is the Modified Bessel Function
of the 2nd kind of order $a$. 

We then  use separation of variables with the definition of the operator  (\ref{eq:D2explanation}) to
solve equation (\ref{eq:D^2*psi_1 Solution}) for $\psi_{1}$ noting that (\ref{eq:D^2*psi_1 Solution})
gives the particular solution of $\psi_{1}$. Hence we finally obtain
\begin{multline}
\psi_{1}=e^{it}\left[\left(\int_{\mu}^{1}P_{0}(x)dx+e_{0}\right)\left(A_{0}\frac{\sqrt{r}}{\alpha^{2}}I_{\frac{1}{2}}(\alpha r)+B_{0}\frac{\sqrt{r}}{\alpha^{2}}K_{\frac{1}{2}}(\alpha r)+C_{0}r+D_{0}\right)+\right.\\
\left.\sum_{n=1}^{\infty}\left(\int_{\mu}^{1}P_{n}(x)dx\right)\left(A_{n}\frac{\sqrt{r}}{\alpha^{2}}I_{n+\frac{1}{2}}(\alpha r)+B_{n}\frac{\sqrt{r}}{\alpha^{2}}K_{n+\frac{1}{2}}(\alpha r)+C_{n}r^{n+1}+D_{n}r^{-n}\right)\right],
\end{multline}
where $\alpha=(1+i)\delta^{-1}$, $I_{a}(x)$ is the Modified Bessel
Function of the 1st kind of order $a$,  and $A_{n}$, $B_{n}$, $C_{n}$, $D_{n}$ are constants to be determined using the boundary conditions. 

\subsection{Enforcing the boundary conditions}

Since $I_{n}(\alpha r)$ increases to infinity exponentially
 as $r\rightarrow\infty$, we first see that the boundary conditions at infinity impose that  $A_{n}=0$ for all $n$ and similarly $C_{n}=0$ for $n>0$. 
 
 Secondly,  singularities in the velocity profile should be removed.  This is not a problem for $u_{r}$ but it is
for $u_{\theta}$. Indeed, we have the scaling $u_{\theta}\sim(1-\mu^{2})^{- \frac{1}{2}}\left(\int_{\mu}^{1}P_{n}(x)dx\right)$, 
which has formal singularities at $\mu=\pm1.$  Using the identity
\begin{equation}
\int_{\mu}^{1}P_{n}(x)dx=\frac{(1-\mu^{2})P_{n}^{'}(\mu)}{n(n+1)}\text{ \,\,\ for }n\neq0,\label{eq:legendreintegraltoderivitive}
\end{equation}
 we see that the singularities are removable for $n>0$ but not for $n=0$.  Since  $e_{0}$ provides only
one degree of freedom, it can be used to remove at most one singularity. Therefore, in order to prevent any singularity in $u_{\theta}$ at $\mu=\pm1$, it is required that $C_{0}=B_{0}=0$. The only term remaining
in $\psi_{1}$ containing $e_{0}$ is proportional to $e_{0}D_{0}$, which is an arbitrary constant and hence
$e_{0}$  can be set to $0$.

With these results, 
 $\psi_{1}$ is reduced 
to 
\begin{equation}
\psi_{1}=e^{it}\left[\left(D_{0}\int_{\mu}^{1}P_{0}(x)dx\right)+\sum_{n=1}^{\infty}\left(\int_{\mu}^{1}P_{n}(x)dx\right)\left(B_{n}\frac{\sqrt{r}}{\alpha^{2}}K_{n+\frac{1}{2}}(\alpha r)+D_{n}r^{-n}\right)\right].\label{eq:FirstOrderSolution}
\end{equation}
Applying the two first-order boundary conditions, (\ref{eq:FirstOrderBoundaryCondition_One})
and (\ref{eq:SecondOrderBoundaryCondition_Two}), finally allows the determination of  the remaining
constants $B_{n}$, $D_{n}$ and $D_{0}$. For ease of expansion later,
a Bessel function factor evaluated at $\alpha$ has been left formally in the
definition of the constants giving us
\begin{align}
B_{n} & =\frac{-(W_{n}+nV_{n})}{K_{n+\frac{1}{2}}(\alpha)}\left(\frac{(1+i)}{\delta}+n+\delta\frac{(i-1)n^{2}}{4}\right)+O\left(\frac{\delta^{2}}{K_{n+\frac{1}{2}}(\alpha)}\right)\:\text{ for }n\ge1,\label{eq:ConstantValueBessel}\\
D_{n} & =V_{n}+\frac{\delta(1-i)(W_{n}+nV_{n})}{2}-\frac{in(W_{n}+nV_{n})\delta^{2}}{2}+O(\delta^{3})\text{ \,\ for }n\ge1,\label{eq:ConstantValueNegativePolynomial}\\
D_{0} & =V_{0}.\label{eq:ConstantValueZero}
\end{align}
For simplicity of notation in what follows,  we define $W_{0}=0$ and $B_{0}=0$ so that \eqref{eq:ConstantValueBessel} and similarly \eqref{eq:ConstantValueNegativePolynomial} remain valid for $n=0$.

\section{Second-order asymptotic solution\label{sec:Second_order_Asymptotic_solution}}

\subsection{Second-order boundary conditions}
\label{sec:Second_Order_Boundary_Conditions}
Equations  (\ref{eq:FirstOrderSolution})-(\ref{eq:ConstantValueZero}) give the full solution for $\psi_{1}$. Using (\ref{eq:TaylorExpansin}) we can now time-average the boundary conditions at order $\epsilon^2$. Note that the product of two terms of the form $fe^{it}$ and $ge^{it}$ time average to $f \overline{g}/2$ or $ \overline{f} g/2$ where the real part is assumed and overbars denote complex conjugates.
Simplifying the  $\mu$ dependence of the result to a sum over the appropriate basis functions, i.e.~$(1-\mu^{2})^{-\frac{1}{2}}{\int_{\mu}^{1}P_{n}(x)dx}$ for $u_{\theta}$ and $P_{n}(\mu)$ for $u_{r}$ 
and using the classical identities 
\begin{eqnarray}
xP_{n}^{'}(x)&=&P_{n+1}^{'}(x)-(n+1)P_{n}(x),\\
P_{n+1}^{'}(x)&=&P_{n-1}^{'}(x)+(2n+1)P_{n}(x),
\end{eqnarray}
we explicitly  obtain time-averaged boundary conditions
\begin{multline}
\langle u_{r}^{(2)} \rangle \bigg\vert_{r=1}=\sum_{k=1}^{\infty}\left\{ \frac{i}{2}\left[2V_{0}\bar{V}_{k}-\sum_{n=0}^{\infty}\sum_{m=1}^{\infty}\left(W_{m}-2V_{m}\right)\bar{V}_{n}g_{knm}\right.\right.\\
\left.\left.+\sum_{n=1}^{\infty}\sum_{m=1}^{\infty}\frac{W_{m}\left(\bar{V}_{n}n(n+1)-\bar{W}_{n}\right)}{mn(n+1)(m+1)}f_{knm}\right]\right\} P_{k}(\mu),\label{eq:RadialCondition_SecondOrderProper}
\end{multline}
and
\begin{multline}
\langle u_{\theta}^{(2)} \rangle \bigg\vert_{r=1}=\sum_{k=1}^{\infty}\left\{ \sum_{n=0}^{\infty}\sum_{m=1}^{\infty}a_{knm}\frac{i}{2}\left[2V_{n}\bar{W}_{m}-\bar{V}_{n}\left(\left(W_{m}+mV_{m}\right)\frac{\left(1+i\right)}{\delta}\right.\right.\right.\\
\left.\left.\left.\phantom{\frac{(1+i)}{\delta}}+\left((m+1)W_{m}-mV_{m}\right)\right)-\bar{W}_{m}\left(W_{n}-\sum_{j=1}^{\infty}\frac{C_{nj}}{j(j+1)}W_{j}\right)\right]\right\} \left(\frac{\int_{\mu}^{1}P_{k}(x)dx}{\left(1-\mu^{2}\right)^{\frac{1}{2}}}\right),\label{eq:AngularCondition_SecondOrderProper}
\end{multline}
where we have used triangular brackets to indicate time-averaging.

In these equations, the series of coefficients $C_{nj}$,  $a_{knm}$, $f_{knm}$ and $g_{knm}$
  are defined by 
\begin{equation}
C_{nj}=\begin{cases}
0 & \text{if }(j<n)\text{ or ($n$ and $j$ have different parity)},\\
n & \text{if }j=n,\\
(2n+1) & \text{otherwise},
\end{cases}\label{eq:C_nj}
\end{equation}
and
\begin{align}
P_{n}(\mu)\left(\int_{\mu}^{1}P_{m}(x)dx\right) & =\sum_{k=1}^{\infty}a_{knm}\left(\int_{\mu}^{1}P_{k}(x)dx\right),\label{eq:a_knm}\\
P_{n}^{1}(\mu)P_{m}^{1}(\mu) & =\sum_{k=0}^{\infty}f_{knm}P_{k}(\mu),\label{eq:f_knm}\\
P_{n}(\mu)P_{m}(\mu) & =\sum_{k=0}^{\infty}g_{knm}P_{k}(\mu),\label{eq:g_knm}
\end{align}
and
where $P_{n}^{1}(\mu)$ is the  Associated Legendre Polynomial
of degree $n$ and order 1. Recall that 
Associated Legendre Polynomials of degree $n$ and
order $m$ are defined by 
\begin{equation}
P_{n}^{m}(x)=(1-x^{2})^{\frac{m}{2}}\frac{d^{m}P_{n}(x)}{dx^{m}},\label{eq:associatedlegendredefinition}
\end{equation}
and have the useful orthogonality property 
\begin{equation}
\int_{-1}^{1}P_{n}^{m}(x)P_{n'}^{m'}(x)dx=\frac{2}{(2n+1)}\frac{(n+m)!}{(n-m)!}\delta_{nn'}\delta_{mm'}.\label{eq:Associatedlengendreorthoganal}
\end{equation}
Using the orthogonality property  together with   \eqref{eq:legendreintegraltoderivitive}, the formulae for $a_{knm}$ (\ref{eq:a_knm}), $f_{knm}$ (\ref{eq:f_knm}) and $g_{knm}$, 
 (\ref{eq:g_knm}) can be rearranged into a
more useful form to calculate their numerical value
\begin{eqnarray}
a_{knm}&=&\frac{k(k+1)}{m(m+1)}\frac{\int_{-1}^{1}P_{n}(x)P_{m}^{1}(x)P_{k}^{1}(x)dx}{\int_{-1}^{1}P_{k}^{1}(x)P_{k}^{1}(x)dx}=\frac{(2k+1)}{2m(m+1)}\int_{-1}^{1}P_{n}(x)P_{m}^{1}(x)P_{k}^{1}(x)dx,\\
f_{knm}&=&\frac{(2k+1)}{2}\int_{-1}^{1}P_{n}^{1}(x)P_{m}^{1}(x)P_{k}(x)dx,\\
g_{knm}&=&\frac{(2k+1)}{2}\int_{-1}^{1}P_{k}(x)P_{n}(x)P_{m}(x)dx.
\end{eqnarray}
$a_{knm}$, $f_{knm}$ and $g_{knm}$ are Gaunt Coefficients, which have been  extensively studied due to their appearance in theoretical physics. Gaunt's formula \cite{Gaunt_1929_TripleAssLeg} for the triple product integral and fast numerical algorithms \cite{FastGaunt_1996} exist to evaluate such coefficients.

\subsection{Nonlinear Forcing}

The  governing equation at order $\epsilon^{2}$ is given by
\begin{equation}
\frac{\delta^{2}}{2}\left(D^{4}\psi_{2}\right)-\frac{\partial(D^{2}\psi_{2})}{\partial t}=\frac{1}{r^{2}}\left(\frac{\partial(\psi_{1},D^{2}\psi_{1})}{\partial(r,\mu)}+2L\psi_{1}D\psi_{1}\right).\label{eq:secondordergoverningequation}
\end{equation}
Time averaging equation (\ref{eq:secondordergoverningequation}) leads to
\begin{equation}
\frac{\delta^{2}}{2}D^{4} \langle \psi_{2} \rangle =\frac{1}{r^{2}} \left \langle\frac{\partial(\psi_{1},D^{2}\psi_{1})}{\partial(r,\mu)}+2L\psi_{1}D\psi_{1} \right \rangle.\label{eq:secondordergoverningequationproper}
\end{equation}

A general second order solution that is valid throughout the domain
cannot be found due to the complexity of the right-hand side of equation \eqref{eq:secondordergoverningequationproper}. However, for small values of $\delta$, solutions can be found separately within the viscous boundary layer and in the far field, and they can be  asymptotically matched to provide a full external solution. This is the method we will be using in this paper.  In either case, an indication of the form of the non-linear
forcing is required. The term $D^{2}\psi_{1}$ is given in equation (\ref{eq:D^2*psi_1 Solution}). We also have 
\begin{multline}
L\psi_{1}=e^{it}\left[\sum_{n=1}^{\infty}\left(\frac{\mu P_{n}^{'}(\mu)}{n(n+1)}\right)\left(\frac{B_{n}}{2\alpha^{2}\sqrt{r}}K_{n+\frac{1}{2}}(\alpha r)+\frac{B_{n}\sqrt{r}}{\alpha}K_{n+\frac{1}{2}}^{'}(\alpha r)-D_{n}nr^{-(n+1)}\right)\right.\\
\left.-\frac{D_{0}P_{0}(\mu)}{r}-\sum_{n=1}^{\infty}\left(P_{n}(\mu)\right)\left(\frac{B_{n}}{\sqrt{r}\alpha^{2}}K_{n+\frac{1}{2}}(\alpha r)+D_{n}r^{-(n+1)}\right)\right],
\label{44}
\end{multline}
and
\begin{multline}
\left \langle \frac{\partial(\psi_{1},D^{2}\psi_{1})}{\partial(r,\mu)} \right \rangle =-\frac{1}{2}\sum_{n=0}^{\infty}\sum_{m=1}^{\infty}P_{n}(\mu)\left(\int_{\mu}^{1}P_{m}(x)dx\right)\left(\frac{\bar{B}_{n}B_{m}}{\alpha^{2}}\bar{K}_{n+\frac{1}{2}}(\alpha r)K_{m+\frac{1}{2}}(\alpha r)\right.\\
+2\frac{\bar{B}_{n}B_{m}r}{\alpha}\bar{K}_{n+\frac{1}{2}}(\alpha r)K_{m+\frac{1}{2}}^{'}(\alpha r)-D_{m}\bar{B}_{n}mr^{-(m+\frac{1}{2})}\bar{K}_{n+\frac{1}{2}}(\alpha r)\\
\left.-\frac{B_{m}\bar{D}_{n}}{2}K_{m+\frac{1}{2}}(\alpha r)r^{-(n+\frac{1}{2})}-B_{m}\bar{D}_{n}\alpha r^{-(n-\frac{1}{2})}K_{m+\frac{1}{2}}^{'}(\alpha r)\right).\label{45}
\end{multline}

Using the coefficient $a_{knm}$ defined in (\ref{eq:a_knm}), equation 
\eqref{45} can then be transformed to the appropriate (integral polynomial) basis. Similarly the quantity given by twice \eqref{44} multiplied by \eqref{eq:D^2*psi_1 Solution} can have its basis transformed using $a_{knm}$ and $C_{nj}$ defined in (\ref{eq:C_nj}).  These two quantities can then be added which gives the total non-linear forcing as
\begin{multline}
\left \langle 2L\psi_{1}D^{2}\psi_{1}+\frac{\partial(\psi_{1},D^{2}\psi_{1})}{\partial(r,\mu)} \right \rangle =\\
\sum_{k=1}^{\infty}\left(\int_{\mu}^{1}P_{k}(x)dx\right)\left[\sum_{n=0}^{\infty}\sum_{m=1}^{\infty}a_{knm}\left(\frac{\bar{B}_{n}B_{m}}{2\alpha^{2}}\bar{K}_{n+\frac{1}{2}}(\alpha r)K_{m+\frac{1}{2}}(\alpha r)\right.\right.\\
-\frac{3}{4}\bar{D}_{n}B_{m}r^{-(n+\frac{1}{2})}K_{m+\frac{1}{2}}(\alpha r)-\frac{\bar{B}_{n}B_{m}r}{\alpha}\bar{K}_{n+\frac{1}{2}}(\alpha r)K_{m+\frac{1}{2}}^{'}(\alpha r)+\frac{1}{2}D_{m}\bar{B}_{n}mr^{-(m+\frac{1}{2})}\bar{K}_{n+\frac{1}{2}}(\alpha r)\\
+\frac{1}{2}B_{m}\bar{D}_{n}\alpha r^{-(n-\frac{1}{2})}K_{m+\frac{1}{2}}^{'}(\alpha r)+\sum_{j=1}^{\infty}\left(\frac{C_{nj}}{j(j+1)}\right)\left(-\frac{\bar{B_{j}}B_{m}}{2\alpha^{2}}\bar{K}_{j+\frac{1}{2}}(\alpha r)K_{m+\frac{1}{2}}(\alpha r)\right.\\
\left.\left.\left.+\frac{\bar{B}_{j}B_{m}r}{\bar{\alpha}}K_{m+\frac{1}{2}}(\alpha r)\bar{K^{'}}_{j+\frac{1}{2}}(\alpha r)-\bar{D}_{j}jr^{-(j+\frac{1}{2})}K_{m+\frac{1}{2}}(\alpha r)B_{m}\right)\right)\right].\label{eq:SecondOrderForcingMostOf}
\end{multline}

\subsection{Solution inside the boundary layer}

The rest boundary is located at $r=1$ with a boundary layer of size $\delta$. We thus define an inner variable $\eta$ related to $r$ by $r=1+\delta\eta$. The
boundary layer is small, $\delta\ll1$, and within the boundary layer
the inner variable $\eta$ varies from $0$ to $1$. We now write the  second-order
equation  in terms of $\eta$ and expand in ascending powers of $\delta$. 

First consider expanding the right-hand side of equation (\ref{eq:secondordergoverningequationproper}), i.e.~equation (\ref{eq:SecondOrderForcingMostOf}) divided by $r^2$.  
When the
first-order boundary conditions were applied above, we obtained in equations  (\ref{eq:ConstantValueBessel})-(\ref{eq:ConstantValueZero}) Taylor expansions of the coefficients $D_{k}$
and $B_{k}$ in terms of $\delta$.   The powers of $r$ can also be Taylor expanded about $r=1$ to also obtain
a powers series in $\delta$.

 However the Taylor expansions of the Bessel Functions have to be done more carefully. A useful identity \cite{BesselPoly_1978} for expanding the Bessel Function is that, for non-negative integer $n$ we have
\begin{equation}
K_{n+\frac{1}{2}}(z)=\sqrt{\frac{\pi}{2}}\frac{e^{-z}}{\sqrt{z}}\sum_{j=0}^{n}\frac{\left(j+n\right)!}{\left(n-j\right)!j!}(2z)^{-j}.
\end{equation}
Next, notice that when the expansions for $D_{k}$ and $B_{k}$ are substituted
into (\ref{eq:secondordergoverningequationproper}), the Bessel functions
 always appear in ratios of the form
\begin{equation}
 \frac{\hat{K}_{n+\frac{1}{2}}(\alpha r)}{K_{n+\frac{1}{2}}(\alpha)},
\end{equation}
where $\hat{K}$ represents a derivative or complex conjugate of
the Bessel Function. As such, taking $f(r)$ as the appropriate power
series in $r$ gives 
\begin{equation}
\frac{\hat{K}_{j+\frac{1}{2}}(\alpha r)}{K_{j+\frac{1}{2}}(\alpha)}=e^{-\tilde{\alpha}(r-1)}f(r)=e^{-(1\pm i)\eta}f(1+\delta\eta),
\end{equation}
where $\tilde{\alpha}$ can be $\alpha$ or its complex conjugate. 
The exponential part of the Bessel function can clearly not be Taylor expanded in powers of $\delta$. Hence to obtain the correct expansion, the power series part of the Bessel Function should be Taylor expanded in powers of $\delta$ and the negative exponential part should only have the appropriate $\eta$ substitution carried out. 

Upon completion of the substitution and expansion in $\delta$, we obtain 
\begin{multline}
\frac{1}{r^{2}}\left \langle 2L\psi_{1}D^{2}\psi_{1}+\frac{\partial(\psi_{1},D^{2}\psi_{1})}{\partial(r,\mu)}\right \rangle =\sum_{k=1}^{\infty}\left(\int_{\mu}^{1}P_{k}(x)dx\right)\sum_{n=0}^{\infty}\sum_{m=1}^{\infty}a_{knm}\times\\
\left\{ \frac{3}{4}(W_{m}+mV_{m})e^{-(1+i)\eta}\left(\frac{\bar{V}_{n}(1+i)}{\delta}\right)+(\bar{W}_{n}+n\bar{V}_{n})(W_{m}+mV_{m})\left(\frac{1-i}{\delta}\right)e^{-2\eta}\phantom{\frac{\bar{V}_{n}(1+i)}{\delta}}\right.\\
-\frac{1}{2}m(\bar{V}_{n}n+\bar{W}_{n})\left(V_{m}\frac{(1-i)}{\delta}\right)e^{-(1-i)\eta}+\frac{1}{2}(W_{m}+mV_{m})\left[\frac{2i\bar{V}_{n}}{\delta^{2}}+\right.\\
\left.\frac{1}{\delta}\left(-2inV_{n}\eta+\bar{V}_{n}m(1+i)-(1-i)(\bar{W}_{n}+n\bar{V}_{n})+(i+1)\frac{\bar{V}_{n}}{2}-4iV_{n}\eta\right)\right]e^{-(1+i)\eta}\\
+\sum_{j=1}^{\infty}\left(\frac{C_{nj}}{j(j+1)}\right)\left[-(\bar{W}_{j}+j\bar{V}_{j})(W_{m}+mV_{m})e^{-2\eta}\left(\frac{(1+i)}{\delta}\right)\phantom{\frac{\bar{V}_{j}j(1+i)}{\delta}}\right.\\
\left.\left.+(W_{m}+mV_{m})e^{-(1+i)\eta}\left(\frac{\bar{V}_{j}j(1+i)}{\delta}\right)\right]+O(1)\right\} .\label{eq:SecondOrderExpansioin}
\end{multline}

Inside the boundary layer, the $D^4$ operator  on the left hand side of (\ref{eq:secondordergoverningequationproper}) is asymptotically given by
\begin{equation}
D^{4}=\frac{1}{\delta^{4}}\frac{\partial^{4}}{\partial\eta^{4}}+O(\delta^{-2})\label{eq:LeadingOrderD4},
\end{equation}
and therefore (\ref{eq:secondordergoverningequationproper}) finally simplifies to\begin{multline}
\left \langle \frac{\partial^{4} \psi_{2}}{\partial\eta^{4}} \right \rangle =\sum_{k=1}^{\infty}\left(\int_{\mu}^{1}P_{k}(x)dx\right)\sum_{n=0}^{\infty}\sum_{m=1}^{\infty}a_{knm}\left\{ \frac{3}{2}\delta(W_{m}+mV_{m})e^{-(1+i)\eta}\bar{V}_{n}(1+i)\right.\\
+2\delta(\bar{W}_{n}+n\bar{V}_{n})(W_{m}+mV_{m})\left(1-i\right)e^{-2\eta}-\delta m(\bar{V}_{n}n+\bar{W}_{n})V_{m}(1-i)e^{-(1-i)\eta}\\
+(W_{m}+mV_{m})\left[\delta\left(-2in\bar{V}_{n}\eta+\bar{V}_{n}m(1+i)-(1-i)(\bar{W}_{n}+n\bar{V}_{n})+(i+1)\frac{\bar{V}_{n}}{2}-4i\bar{V}_{n}\eta\right)\right.\\
\left.\phantom{\frac{\bar{V}_{n}}{2}}+2i\bar{V}_{n}\right]e^{-(1+i)\eta}+\sum_{j=1}^{\infty}\left(\frac{C_{nj}}{j(j+1)}\right)\left[-2\delta(\bar{W}_{j}+j\bar{V}_{j})(W_{m}+mV_{m})e^{-2\eta}\left(1+i\right)\right.\\
\left.\left.+2\delta(W_{m}+mV_{m})e^{-(1+i)\eta}\left(\bar{V}_{j}j(1+i)\right)\right]\right\} +O(1).\label{eq:IntegralEquation}
\end{multline}

Using the elementary integrals
\begin{equation}
\int\!\!\!
\int\!\!\!
\int\!\!\!
\int
 e^{ax}dx=\frac{e^{ax}}{a^{4}},\quad
 \int\!\!\!
\int\!\!\!
\int\!\!\!
\int xe^{ax}dx=\frac{xe^{ax}}{a^{4}}-\frac{4e^{ax}}{a^{5}},
\end{equation}
allows us to integrate equation (\ref{eq:IntegralEquation}) explicitly, leading to the  general inner solution 
\begin{multline}
\left \langle \psi_{2} \right \rangle =\sum_{k=1}^{\infty}\left\{ \left(\int_{\mu}^{1}P_{k}(x)dx\right)\sum_{n=0}^{\infty}\sum_{m=1}^{\infty}a_{knm}\left[-\frac{3}{8}(W_{m}+mV_{m})e^{-(1+i)\eta}\bar{V}_{n}(1+i)\delta\right.\right.\\
+\delta(\bar{W}_{n}+n\bar{V}_{n})(W_{m}+mV_{m})\frac{(1-i)}{8}e^{-2\eta}+m(\bar{V}_{n}n+\bar{W}_{n})V_{m}\frac{(1-i)}{4}\delta e^{-(1-i)\eta}\\
-(W_{m}+mV_{m})\left\{ \frac{\delta}{4}\left[-2i(n+2)\bar{V}_{n}\left(\eta+2(1-i)\right)+\bar{V}_{n}m(1+i)-(1-i)(\bar{W}_{n}+n\bar{V}_{n})\right.\right.\\
\left.\left.+(i+1)\frac{\bar{V}_{n}}{2}\right]+\frac{i\bar{V}_{n}}{2}\right\} e^{-(1+i)\eta}+\sum_{j=1}^{\infty}\left(\frac{C_{nj}}{j(j+1)}\right)\left(-(\bar{W}_{j}+j\bar{V}_{j})(W_{m}+mV_{m})e^{-2\eta}\frac{(1+i)}{8}\delta\phantom{\frac{\bar{V}_{j}}{2}}\right.\\
\left.\left.\left.-\delta(W_{m}+mV_{m})\frac{e^{-(1+i)\eta}}{2}\bar{V}_{j}j(1+i)\right)+O(\delta^{2})\right]+(L_{k}+M_{k}\eta+N_{k}\eta^{2}+Q_{k}\eta^{3})\right\} ,\label{eq:InnerSolution}
\end{multline}
where $L_{k}$, $M_{k}$, $N_{k}$ and $Q_{k}$ are constants of integration
to be determined. Of those, two will be determined by enforcing the 
two second-order boundary conditions (\ref{eq:RadialCondition_SecondOrderProper})
and (\ref{eq:AngularCondition_SecondOrderProper}), namely
$L_{k}$ and $M_{k}$.  Since these boundary conditions must hold for all $-1\leq \mu\leq 1$, the coefficient of each basis function (i.e.~$P_{n}(\mu)$
or $\int_{\mu}^{1}P_{n}(x)dx$) must obey these boundary conditions term by term,
giving a countably infinite number of equations with solution
\begin{eqnarray}
L_{k}&=&\Re\left\{ \frac{i}{2}\left[\sum_{n=0}^{\infty}\sum_{m=1}^{\infty}a_{knm}(W_{m}+mV_{m})\bar{V}_{n}+2V_{0}\bar{V}_{k}-\sum_{n=0}^{\infty}\sum_{m=1}^{\infty}\left(W_{m}-2V_{m}\right)\bar{V}_{n}g_{knm}\right.\right.\notag\\
&&\quad\left.\left.+\sum_{n=1}^{\infty}\sum_{m=1}^{\infty}\frac{W_{m}\left(\bar{V}_{n}n(n+1)-\bar{W}_{n}\right)}{mn(n+1)(m+1)}f_{knm}\right]\right\} +O(\delta),\quad k>0,\label{eq:Lt-1}\\
M_{k}&=&\delta\left(\sum_{n=0}^{\infty}\sum_{m=1}^{\infty}a_{knm}\left\{ \frac{-imV_{m}(\bar{W}_{n}+n\bar{V}_{n})}{2}+\frac{i\bar{V_{n}}}{2}(W_{m}+mV_{m})(4+3n-m)\right.\right.\notag \\
\quad \notag &&+\frac{(3-i)}{4}(\bar{W}_{n}+n\bar{V}_{n})(W_{m}+mV_{m})+(m+3)\frac{i}{2}\bar{V}_{n}W_{m}-m\frac{i}{2}\bar{V}_{n}V_{m}\\
\quad &&+\frac{i}{2}\bar{W}_{m}W_{n}-\sum_{j=1}^{\infty}\left(\frac{C_{nj}}{j(j+1)}\right)\left[(\bar{W}_{j}+j\bar{V}_{j})(W_{m}+mV_{m})\left(\frac{1+i}{4}\right)\right.\notag \\
\quad &&\left.\left.\left.+ij\bar{V}_{j}(W_{m}+mV_{m})+\frac{i}{2}\bar{W}_{m}W_{j}\right]\right\} \right)+O(\delta^{2}),\quad k>0.\label{eq:Mt}
\end{eqnarray}
Asymptotic matching will then determine the values of the remaining coefficients. 

\subsection{Solution Outside the Boundary Layer}
Looking at equation (\ref{eq:SecondOrderForcingMostOf}), we see that all the terms in the non-linear forcing are multiples of Modified Bessel Functions
of the second kind or their derivatives. As such, this forcing decays
away exponentially fast as $r\rightarrow\infty$ and can be neglected outside the boundary layer. Therefore in this region we have a Stokes flow and the governing equation is
\begin{equation}
D^{2} D^{2}\langle \psi_{2} \rangle = 0,\label{eq:outersecondordergoverningequation}
\end{equation}
with exponentially small errors. 

Given the  form of the inner solution \eqref{eq:InnerSolution}, and anticipating the asymptotic matching, we can look for the outer solution with a known $\mu$ dependence 
as
\begin{equation}
\langle \psi_{2} \rangle=\sum_{n=1}^{\infty}f_{n}(r)\left(\int_{\mu}^{1}P_{n}(x)dx\right)+f_{0}(r)\left(\int_{\mu}^{1}P_{0}(x)dx+e_{0}\right).\label{eq:secondorderformofsolution}
\end{equation}
The value of $ \langle D^{2} \psi_{2} \rangle $ can be found by differentiating (\ref{eq:secondorderformofsolution})
but also by solving (\ref{eq:outersecondordergoverningequation})
for $ \langle D^{2} \psi_{2} \rangle$ using separation of variables. Equating
these gives a second order differential equation for $f$ with power-law solutions. 
The general outer solution is thus given by
\begin{eqnarray}
\langle \psi_{2} \rangle &=&\left(R_{0}+T_{0}r+Y_{0}r^{2}+S_{0}r^{3}\right)\left(\int_{\mu}^{1}P_{0}(x)dx+e_{0}\right)\notag \\
&&+\sum_{n=1}^{\infty}\left(R_{n}r^{n+3}+T_{n}r^{-n}+Y_{n}r^{n+1}+S_{n}r^{-(n-2)}\right)\left(\int_{\mu}^{1}P_{n}(x)dx\right).\label{eq:GeneralOuterFormulation}
\end{eqnarray}
Applying the boundary condition at infinity 
gives $R_{n}=Y_{n}=0$ and $S_{0}=Y_{0}=0$. Furthermore, in order to avoid a singularity
in $u_{\theta}$ at $\mu=\pm1$ it is required $T_{0}=0,$ and as
$\bar{\psi}_{2}$ is a stream function it can be set that $e_{0}=0$
without affecting $u_{r}$ or $u_{\theta}$. Hence we obtain the outer solution as
\begin{equation}
\langle \psi_{2} \rangle =R_{0}\left(\int_{\mu}^{1}P_{0}(x)dx\right)+\sum_{n=1}^{\infty}\left(T_{n}r^{-n}+S_{n}r^{-(n-2)}\right)\left(\int_{\mu}^{1}P_{n}(x)dx\right).\label{eq:OuterSolution}
\end{equation}

\subsection{Matching}

The final part of determining the solution for the flow consists of carrying  out the asymptotic matching between the inner (\ref{eq:InnerSolution}) and outer solutions \eqref{eq:OuterSolution}. We first need to evaluate the inner solution, (\ref{eq:InnerSolution}), in the limit $\eta\gg1$ which, because of the  negative exponentials, simplifies to
\begin{equation}
\left \langle \psi_{2} \right \rangle =\sum_{k=1}^{\infty}\left(\int_{\mu}^{1}P_{k}(x)dx\right)\left(L_{k}+M_{k}\eta+N_{k}\eta^{2}+Q_{k}\eta^{3}\right).
\end{equation}
The outer solution, (\ref{eq:OuterSolution}), then needs to be evaluated 
in the limit $r\to 1$. Writing the outer solution
in terms of the inner variable $\eta$ and  Taylor expanding the expression about $\eta=0$
then gives 
 \begin{multline}
\left \langle \psi_{2} \right \rangle =R_{0}\left(\int_{\mu}^{1}P_{0}(x)dx\right)+\sum_{n=1}^{\infty}\left(\int_{\mu}^{1}P_{n}(x)dx\right)\left[(T_{n}+S_{n})-(nT_{n}+(n-2)S_{n})\delta\eta\phantom{\frac{1}{2}}\right.\\
+\left(\frac{n(n+1)}{2}T_{n}+\frac{(n-2)(n-1)}{2}S_{n}\right)\delta^{2}\eta^{2}\\
\left.-\left(\frac{n(n+1)(n+2)}{6}T_{n}+\frac{n(n-1)(n-2)}{6}S_{n}\right)\delta^{3}\eta^{3}
+O(\eta^{4})\right].
\end{multline}
Equating the two highest orders of $\eta$ gives
\begin{eqnarray}
L_{k}&=&T_{k}+S_{k}\quad  (k\ge1),\label{eq:L_k to T_k}\\
M_{k}&=&-\left[kT_{k}+(k-2)S_{k}\right]\delta \quad  (k\ge1),\label{eq:M_k to T_k}\\
R_{0}&=& 0   \label{eq:R_0 to L_0}.
\end{eqnarray}
If we use $M_{k}^{(\delta)}$ to denote the  $O(\delta)$
term of $M_{k}$, the  outer constants are thus given by
\begin{eqnarray}
T_{k} & = & \frac{(2-k)}{2}L_{k}^{(1)}-\frac{M_{k}^{(\delta)}}{2\delta}+O(\delta)\quad  (k\ge1),\label{eq:T_K to M_k value}\\
S_{k} & = & \frac{k}{2}L_{k}^{(1)}+\frac{M_{k}^{(\delta)}}{2\delta}+O(\delta)\quad (k\ge1).\label{eq:S_k to M_k value}
\end{eqnarray}
Notice that $T_{k}$ and $S_{k}$ are now known so they can be used
to determine $N_{k}$ and $Q_{k}$ by matching to third and fourth
order; if one carries out this matching, one obtains that $N_{k}$ and $Q_{k}$ are  $O(\delta^{2})$
and $O(\delta^{3}$) respectively.

\section{Lagrangian streaming\label{sec:LagrangianStreaming} }
The solution derived so far has focused on the  Eulerian streaming, i.e.~the time-averaged Eulerian velocity field at a fixed position in the laboratory frame of reference.  In order to compare with future   experimental results tracking the motion of passive tracers in the flow, it is necessary to calculate the Lagrangian streaming instead. The difference between the Eulerian and the Lagrangian streaming is the so-called Stokes drift which arises because the Lagrangian particles are advected by the Eulerian velocities at all the positions the particles move through, and not  just fixed positions in the laboratory frame, and thus velocity gradients need to be accounted for.

Longuet-Higgins \cite{Higgins1953_WaterWaves,Higgins1998_Bubble}
showed that the streamfunction for the non-dimensional time averaged Stokes Drift $\bar{\varphi}_{S}$ at   $O(\epsilon^2)$ is given by
\begin{eqnarray}
\langle \varphi_{S} \rangle & = & \overline{\frac{1}{r^{2}}\int\frac{\partial\psi_{1}}{\partial r}dt\frac{\partial\psi_{1}}{\partial\mu}}\cdot
\end{eqnarray}
Ignoring  exponentially-decaying terms, the external solution for the Stokes Drift is thus
\begin{equation}
\langle \varphi_{S} \rangle =-\frac{\epsilon^{2}i}{2\omega}\sum_{k=1}^{\infty}\left(\int_{\mu}^{1}P_{k}(x)dx\right)\left(\sum_{n=0}^{\infty}\sum_{m=1}^{\infty}a_{knm}m\bar{V}_{n}V_{m}r^{-(n+m+3)}\right)+O(\delta).
\end{equation}
Adding this expression to the time-averaged Eulerian solution from equation \eqref{eq:OuterSolution}
leads to the  final expression for the external leading order time averaged Lagrangian streaming as 
\begin{equation}
\langle \psi_{L} \rangle = 
\sum_{k=1}^{\infty}\left(T_{k}r^{-k}+S_{k}r^{-(k-2)}-\sum_{n=0}^{\infty}\sum_{m=1}^{\infty}Y_{knm}r^{-(n+m+3)}\right)\left(\int_{\mu}^{1}P_{k}(x)dx\right),\label{eq:FinalSolutionEquation}
\end{equation}
where 
\begin{equation}
Y_{knm}=\Re\left(\frac{ia_{knm}m\bar{V}_{n}V_{m}}{2}\right),\label{eq:Y_knm}
\end{equation}
and the coefficients $S_{k}$ and $T_{k}$  are given by
\begin{multline}
S_{k}=\Re\left\{ \sum_{n=0}^{\infty}\sum_{m=1}^{\infty}\left(a_{knm}\left\{ \frac{i(W_{m}+mV_{m})k\bar{V}_{n}}{4}\right.\right.\right.\\
-\frac{imV_{m}(\bar{W}_{n}+n\bar{V}_{n})}{4}+\frac{i\bar{V_{n}}}{4}(W_{m}+mV_{m})(4+3n-m)\\
+\frac{(3-i)}{8}(\bar{W}_{n}+n\bar{V}_{n})(W_{m}+mV_{m})+(m+3)\frac{i}{4}\bar{V}_{n}W_{m}-m\frac{i}{4}\bar{V}_{n}V_{m}\\
+\frac{i}{4}\bar{W}_{m}W_{n}-\sum_{j=1}^{\infty}\left(\frac{C_{nj}}{j(j+1)}\right)\left[(\bar{W}_{j}+j\bar{V}_{j})(W_{m}+mV_{m})\left(\frac{1+i}{8}\right)\right.\\
\left.\left.\left.+\frac{i}{2}j\bar{V}_{j}(W_{m}+mV_{m})+\frac{i}{4}\bar{W}_{m}W_{j}\right]\right\} -\frac{ik}{4}\left(W_{m}-2V_{m}\right)\bar{V}_{n}g_{knm}\right)\\
\left.+\sum_{n=1}^{\infty}\sum_{m=1}^{\infty}\left[\frac{ik}{4}\frac{W_{m}\left(\bar{V}_{n}n(n+1)-\bar{W}_{n}\right)}{mn(n+1)(m+1)}f_{knm}\right]+kV_{0}\bar{V}_{t}\frac{i}{2}\right\} +O(\delta),\label{eq:St}
\end{multline}
and
\begin{multline}
T_{k}=\Re\left\{ \sum_{n=0}^{\infty}\sum_{m=1}^{\infty}\left(a_{knm}\left\{ (W_{m}+mV_{m})(2-k)\frac{i}{4}\bar{V}_{n}\right.\right.\right.\\
+\frac{imV_{m}(\bar{W}_{n}+n\bar{V}_{n})}{4}-\frac{i\bar{V_{n}}}{4}(W_{m}+mV_{m})(4+3n-m)\\
-\frac{(3-i)}{8}(\bar{W}_{n}+n\bar{V}_{n})(W_{m}+mV_{m})-(m+3)\frac{i}{4}\bar{V}_{n}W_{m}+m\frac{i}{4}\bar{V}_{n}V_{m}\\
-\frac{i}{4}\bar{W}_{m}W_{n}+\sum_{j=1}^{\infty}\left(\frac{C_{nj}}{j(j+1)}\right)\left[(\bar{W}_{j}+j\bar{V}_{j})(W_{m}+mV_{m})\left(\frac{1+i}{8}\right)\right.\\
\left.\left.\left.+\frac{i}{2}j\bar{V}_{j}(W_{m}+mV_{m})+\frac{i}{4}\bar{W}_{m}W_{j}\right]\right\} -\frac{i}{4}(2-k)\left(W_{m}-2V_{m}\right)\bar{V}_{n}g_{knm}\right)\\
\left.+\sum_{n=1}^{\infty}\sum_{m=1}^{\infty}\left[\frac{i}{4}(2-k)\frac{W_{m}\left(\bar{V}_{n}n(n+1)-\bar{W}_{n}\right)}{mn(n+1)(m+1)}f_{knm}\right]+\frac{i}{2}(2-k)V_{0}\bar{V}_{k}\right\} +O(\delta),\label{eq:Tt}
\end{multline}

We note that in the far field, the flow will be dominated by the slowest spatially
decaying term. In the streamfunction (\ref{eq:FinalSolutionEquation}) this
is the $S_{1}r$ term. The velocities associated with this term decay as $\sim 1/r$ and 
are associated with  a net force acting on the fluid  (Stokeslet). This will be further discussed \S \ref{part:ForceAndVelocity} in the context of force generation and propulsion. 

\section{Special case: Squirming\label{part:SPECIALCASE_Squirmer}}

As discussed in the introduction, the squirmer model of low-Reynolds number swimming  is a popular mathematical model to address the motion of nearly spherical ciliated cells
(e.g.~Opalina, Volvox) \cite{Lighthill1951_Squirmer,Blake1971}. 
The array of deforming cilia  is  modeled as a continuous envelope   where the effective tangential component is deemed far more significant
than the radial component of motion. As such, the position of the surface
can be modeled as fixed at its average position, and imposing steady  velocities along it. Models of this form have been used to study nutrient uptake by microorganisms \cite{Pedley_NutrientUnsteadySquirmer_2005}, interaction of microorganisms \cite{Pedley_TwoSwimmers_2006}, optimal locomotion  \cite{Michelin2010_Squirmer} and were recently generalised \cite{Lauga2014_GeneralisedSqurimingMotion,Ishimoto2014_EffciencyBeyondLighthill}.  Some versions of the squirming model do also allow a radial velocity  to be applied through the fixed boundary as  a model for a porous surface with normal
jets of fluid through it  \cite{Wu_PorousMicroorganism_1977}.

The squirming approximation significantly increases the ease of theoretical
calculations. However, and as expected, it has its limitations. A prescribed forcing through the boundary cannot be used to accurately model a moving boundary. If there is a non-zero radial velocity at the fixed boundary, the streaming flow will become one order of magnitude larger since the boundary conditions are no longer canceling the leading-order term. In  the very specific case of solely radial motion where all the radial modes are exactly ${\pi}/{4}$ out of phase with
each other the solution will be at the right order, but other  important terms will still be missing from the streaming. This demonstrates the physical movement
of the boundary, and hence the physical displacement of the fluid in that
region, is as important as the prescribed velocities it is imparting
to the fluid around it.

For angular motion alone, however, the squirmer streaming is identical to the full solution, demonstrating such an approximation is valid. Furthermore, in that situation  the Lagrangian and Eulerian streamings are  the same since the Stokes Drift   depends only on the radial motion of the surface. We thus focus on the standard  tangential squirmer model where there is no radial  motion at leading-order and  $V_n=0$ for all $n$. In that setup, the  generated streaming is of the same form as has already been derived, (\ref{eq:FinalSolutionEquation}), but the  constants are much simpler with $Y_{knm}=0$ and 
\begin{multline}
S_{k}=\Re\left\{ \sum_{n=0}^{\infty}\sum_{m=1}^{\infty}a_{knm}\left[\frac{3(1-i)}{8}\bar{W}_{n}W_{m}-\sum_{j=1}^{\infty}\left(\frac{C_{nj}}{j(j+1)}\right)\bar{W}_{j}W_{m}\left(\frac{1-i}{8}\right)\right]\right.\\
\left.+\sum_{n=1}^{\infty}\sum_{m=1}^{\infty}\left[-\frac{ik}{4}\frac{W_{m}\bar{W}_{n}}{mn(n+1)(m+1)}f_{knm}\right]\right\} +O(\delta),
\end{multline}
and
\begin{multline}
T_{k}=\Re\left\{ \sum_{n=0}^{\infty}\sum_{m=1}^{\infty}a_{knm}\left[-\frac{3(1-i)}{8}\bar{W}_{n}W_{m}+\sum_{j=1}^{\infty}\left(\frac{C_{nj}}{j(j+1)}\right)\bar{W}_{j}W_{m}\left(\frac{1-i}{8}\right)\right]\right.\\
\left.+\sum_{n=1}^{\infty}\sum_{m=1}^{\infty}\left[-\frac{i}{4}(2-k)\frac{W_{m}\bar{W}_{n}}{mn(n+1)(m+1)}f_{knm}\right]\right\} +O(\delta).
\end{multline}

\section{Allowing Slip on the Boundary\label{sec:AllowingBoundarySlip}}

In previous sections we assumed that the fluid satisfied the  no-slip boundary condition and thus  exactly matched the motion of our deforming body. 
Slip can however  be systematically incorporated into this model through a small change of boundary
conditions. The general form of the solution  remains the same but the
constants of integration  gain an extra contribution. This will
then extend the model to include other spherical bodies, in particular   bubbles. 

In the case of  no-slip, the motion of the boundary was described by its radial
position, $R$, and angular position, $\Theta$. However instead $R$
and $\Theta$ can be interpreted as describing the motion of the fluid on the
boundary of the spherical body. If we allow streaming at second
order on the boundary, we can then  write 
\begin{eqnarray}
R&=&1-\epsilon\sum_{n=0}^{\infty}V_{n}P_{n}(\mu)e^{i(t+\frac{\pi}{2})}+\epsilon^{2}\sum_{n=0}^{\infty}G_{n}P_{n}(\mu)g(t)+O(\epsilon^{3}),\label{eq:RadialPosition-1}\\
\Theta&=&\theta+\epsilon\sum_{n=1}^{\infty}W_{n}\left(\frac{\int_{\mu}^{1}P_{n}(x)dx}{(1-\mu^{2})^{\frac{1}{2}}}\right)e^{i(t+\frac{\pi}{2})}+\epsilon^{2}\sum_{n=0}^{\infty}F_{n}\left(\frac{\int_{\mu}^{1}P_{n}(x)dx}{(1-\mu^{2})^{\frac{1}{2}}}\right)f(t)+O(\epsilon^{3}),\label{eq:AngularPosition-1}
\end{eqnarray}
which now has the extra second-order contributions where $F_{n}$
and $G_{n}$ are known constants determined by the motion of the spherical
body and $f(t)$ and $g(t)$ are unknown functions of time. 

The new definition of $R$ is equivalent to the previous one since
the fluid and spherical body cannot encompass the same space. Therefore
$V_{n}$ is still determined by the shape of the surface oscillation
and there is still no net motion of the boundary, we thus have $G_{n}=0$ for all $n$. This new definition of $\Theta$
does allow net motion of the fluid along the body's surface, and the coefficients
$W_{n}$ are then chosen so the appropriate surface boundary condition is
obeyed at first order. 
Similarly the value of $F_{n}$ is determined by
ensuring that this same surface boundary condition is obeyed at second
order.  As such $F_{n}$ may be nonzero and without loss of generality we assume that ${\partial f}/{\partial t}$ time averages to one.

The addition of the coefficients  $F_{n}$ is a second-order contribution so $\psi_{1}$ is unchanged from (\ref{eq:FirstOrderSolution}). The form of the second-order inner solution (\ref{eq:InnerSolution}) is unchanged with $L_{k}$ as
in (\ref{eq:Lt-1}) but $M_{k}$ now has an extra $F_{k}$ contribution so we obtain the revised equation
\begin{multline}
M_{k}=\delta\left(\sum_{n=0}^{\infty}\sum_{m=1}^{\infty}a_{knm}\left\{ -i\frac{m}{2}V_{m}(\bar{W}_{n}+n\bar{V}_{n})+\frac{i\bar{V_{n}}}{2}(W_{m}+mV_{m})(4+3n-m)\right.\right.\\
+\frac{(3-i)}{4}(\bar{W}_{n}+n\bar{V}_{n})(W_{m}+mV_{m})+(m+3)\frac{i}{2}\bar{V}_{n}W_{m}-m\frac{i}{2}\bar{V}_{n}V_{m}\\
+\frac{i}{2}\bar{W}_{m}W_{n}-\sum_{j=1}^{\infty}\left(\frac{C_{nj}}{j(j+1)}\right)\left[(\bar{W}_{j}+j\bar{V}_{j})(W_{m}+mV_{m})\left(\frac{1+i}{4}\right)\right.\\
\left.\left.\left.+ij\bar{V}_{j}(W_{m}+mV_{m})+\frac{i}{2}\bar{W}_{m}W_{j}\right]\right\} -F_{k}\right)+O(\delta^{2})\:\text{ for }k>0.\label{eq:Mt-1}
\end{multline}
The new boundary condition considered, such as  that of no angular stress along the surface of a bubble,  would now be applied to this new second order inner solution allowing us to determine the value of the coefficients $F_{k}$.

Similarly the form of the outer solution (\ref{eq:FinalSolutionEquation})
remains the same but through the asymptotic matching there is an extra
contribution to its constants of integration so that the revised formulae for the $S_k$ and $T_k$ coefficients are  now
\begin{multline}
S_{k}=\Re\left\{ \sum_{n=0}^{\infty}\sum_{m=1}^{\infty}\left[a_{knm}\left\{ \frac{i(W_{m}+mV_{m})k\bar{V}_{n}}{4}\right.\right.\right.\\
-\frac{im}{4}V_{m}(\bar{W}_{n}+n\bar{V}_{n})+\frac{i\bar{V_{n}}}{4}(W_{m}+mV_{m})(4+3n-m)\\
+\frac{(3-i)}{8}(\bar{W}_{n}+n\bar{V}_{n})(W_{m}+mV_{m})+(m+3)\frac{i}{4}\bar{V}_{n}W_{m}-m\frac{i}{4}\bar{V}_{n}V_{m}\\
+\frac{i}{4}\bar{W}_{m}W_{n}-\sum_{j=1}^{\infty}\left(\frac{C_{nj}}{j(j+1)}\right)\left[(\bar{W}_{j}+j\bar{V}_{j})(W_{m}+mV_{m})\left(\frac{1+i}{8}\right)\right.\\
\left.\left.\left.+\frac{i}{2}j\bar{V}_{j}(W_{m}+mV_{m})+\frac{i}{4}\bar{W}_{m}W_{j}\right]\right\} -\frac{ik}{4}\left(W_{m}-2V_{m}\right)\bar{V}_{n}g_{knm}\right]\\
\left.+\sum_{n=1}^{\infty}\sum_{m=1}^{\infty}\left[\frac{ik}{4}\frac{W_{m}\left(\bar{V}_{n}n(n+1)-\bar{W}_{n}\right)}{mn(n+1)(m+1)}f_{knm}\right]+kV_{0}\bar{V}_{k}\frac{i}{2}-\frac{F_{k}}{2}\right\} +O(\delta)\label{eq:St-1},
\end{multline}
and
\begin{multline}
T_{k}=\Re\left\{ \sum_{n=0}^{\infty}\sum_{m=1}^{\infty}\left[a_{knm}\left\{ (W_{m}+mV_{m})(2-k)\frac{i}{4}\bar{V}_{n}\right.\right.\right.\\
+\frac{im}{4}V_{m}(\bar{W}_{n}+n\bar{V}_{n})-\frac{i\bar{V_{n}}}{4}(W_{m}+mV_{m})(4+3n-m)\\
-\frac{(3-i)}{8}(\bar{W}_{n}+n\bar{V}_{n})(W_{m}+mV_{m})-(m+3)\frac{i}{4}\bar{V}_{n}W_{m}+m\frac{i}{4}\bar{V}_{n}V_{m}\\
-\frac{i}{4}\bar{W}_{m}W_{n}+\sum_{j=1}^{\infty}\left(\frac{C_{nj}}{j(j+1)}\right)\left[(\bar{W}_{j}+j\bar{V}_{j})(W_{m}+mV_{m})\left(\frac{1+i}{8}\right)\right.\\
\left.\left.\left.+\frac{i}{2}j\bar{V}_{j}(W_{m}+mV_{m})+\frac{i}{4}\bar{W}_{m}W_{j}\right]\right\} -\frac{i}{4}(2-k)\left(W_{m}-2V_{m}\right)\bar{V}_{n}g_{knm}\right]\\
\left.+\sum_{n=1}^{\infty}\sum_{m=1}^{\infty}\left[\frac{i}{4}(2-k)\frac{W_{m}\left(\bar{V}_{n}n(n+1)-\bar{W}_{n}\right)}{mn(n+1)(m+1)}f_{knm}\right]+\frac{i}{2}(2-k)V_{0}\bar{V}_{k}+\frac{F_{k}}{2}\right\} +O(\delta),\label{eq:Tt-1}
\end{multline}
with the coefficients $Y_{knm}$ remaining the same as in (\ref{eq:Y_knm}). Naturally, through the asymptotic matching outlined above, as $T_{k}$ and $S_{k}$ have an extra contribution, the inner constants $N_{k}$ and $Q_{k}$ will also include a contribution proportional to the constants $F_{k}$.

\section{Special case: Free surfaces\label{part:SPECIALCASE_bubble}}

In the case of  an oscillating bubble, the extra boundary condition is that of no 
tangential stress on the bubble surface 
\begin{equation}
\boldsymbol{n}\cdot\boldsymbol{\sigma} \cdot\boldsymbol{t}=0,\quad\text{at }r=R,\,\theta=\Theta,
\end{equation}
 where $\boldsymbol{t}$ is the tangent vector in the plane through
the axis of axisymmetry, $\boldsymbol{n}$ is the normal vector and
$\boldsymbol{\sigma}$ is the Newtonian stress tensor. There is still no penetration
on the bubble surface and the shape of the bubble oscillation (via the coefficients $V_{n}$) is prescribed. Ensuring that the no-tangential stress conditions holds at first order and at  second order (when time averaged) determines the values of 
$W_{n}$ and $F_{n}$. This calculation is quite involved  and its details are given in  Appendices \ref{chap: AppendixA}
and \ref{sec:Appendix2_InPhaseBubble} with the main results quoted in what follows.

\subsection{General case}

In this subsection we calculate the  leading-order streaming provided the result is non-zero (see below).  The generated streaming is of the same form as the one  already derived,
(\ref{eq:FinalSolutionEquation}), with $Y_{knm}$ defined as in (\ref{eq:Y_knm})
and $R_{0}=0$, with the difference that the constants  $T_{k}$ and $S_{k}$ now take new values
\begin{multline}
 T_{k}=\Re\left\{ \frac{(1-k^{2})}{(2k+1)}iV_{0}\bar{V}_{k}+\sum_{n=1}^{\infty}\sum_{m=1}^{\infty}\frac{\left(k^{2}-k-1\right)(n+2)}{2(2k+1)(n+1)(m+1)}f_{knm}i\bar{V}_{n}V_{m}\right.\\
+\sum_{n=0}^{\infty}\sum_{m=1}^{\infty}\frac{3(1-k^{2})}{2(2k+1)}iV_{m}\bar{V}_{n}g_{knm}+\sum_{n=0}^{\infty}\sum_{m=1}^{\infty}\frac{a_{knm}}{2(2k+1)}\left[m\left(n^{2}-4nm-4n\right.\phantom{\frac{c_{nj}}{(}}\right.\\
\left.\left.\left.+m^{2}-m-3\right)i\bar{V}_{n}V_{m}+\sum_{j=1}^{\infty}\left(\frac{C_{nj}}{(j+1)}\right)\left(m(3m+5)iV_{m}\bar{V}_{j}\right)\right]\right\} +O(\delta),\label{eq:T_k_bubble-1}
\end{multline}
and
\begin{multline}
S_{k}=\Re\left\{ \frac{k(k+2)}{(2k+1)}iV_{0}\bar{V}_{k}-\sum_{n=1}^{\infty}\sum_{m=1}^{\infty}\frac{k(n+2)\left(2k+4\right)}{4(n+1)(m+1)\left(2k+1\right)}f_{knm}i\bar{V}_{n}V_{m}\right.\\
+\sum_{n=0}^{\infty}\sum_{m=1}^{\infty}\frac{3k(k+2)}{2(2k+1)}iV_{m}\bar{V}_{n}g_{knm}-\sum_{n=0}^{\infty}\sum_{m=1}^{\infty}\frac{a_{knm}}{2(2k+1)}\left[m\left(n^{2}-4nm-4n\right.\phantom{\frac{c_{nj}}{(}}\right.\\
\left.\left.\left.+m^{2}-m-3\right)i\bar{V}_{n}V_{m}+\sum_{j=1}^{\infty}\left(\frac{C_{nj}}{(j+1)}\right)\left(m(3m+5)iV_{m}\bar{V}_{j}\right)\right]\right\} +O(\delta).\label{eq:S_k_bubble-1}
\end{multline}
This solution is derived in detail in Appendix \ref{chap: AppendixA}.

\subsection{Special case - In-phase motion}

If the $V_n$ coefficients are chosen such that  the real part of $iV_{n}\bar{V}_{m}=0$ for all valued of $ n$ and $m$ (i.e. all modes are in phase or $\pi$ out-of-phase with each other) then $T_{k}=S_{k}=Y_{knm}=0$ and the steady streaming is identically zero at $O(1)$. This of course includes the case where only one mode is being forced. The net streaming in that case occurs at order $O(\delta)$. In order to determine this streaming the solution derived in \S \ref{part:GeneralAcousticStreaming} needs
to be taken to third order, to give one more power of $\delta$, and
then have the no-stress condition applied to it. The details for this calculation
are in Appendix \ref{sec:Appendix2_InPhaseBubble}. The generated
streaming is still of the same form as has already been derived, (\ref{eq:FinalSolutionEquation})
but with $Y_{knm}=0$, $R_{0}=0$ and with the  constants $T_{k}$ and $S_{k}$
now taking new values as
\begin{multline}
T_{k}=\delta\Re\left\{ \sum_{n=1}^{\infty}\sum_{m=1}^{\infty}\frac{(1-k^{2})(n+2)}{2(2k+1)(n+1)}\left(\bar{V}_{n}V_{m}\right)f_{knm}-\sum_{n=0}^{\infty}\sum_{m=0}^{\infty}\frac{(1-k^{2})m(m+2)}{2(2k+1)}\left(V_{m}\bar{V}_{n}\right)g_{knm}\right.\\
+\sum_{n=0}^{\infty}\sum_{m=1}^{\infty}a_{knm}\left[\left(\frac{n^{2}}{4}+\frac{9n}{4}-\frac{5m^{2}}{4}-\frac{5m}{4}+1+\frac{1}{2}k^{2}+\frac{3}{2}k\right)\frac{m(m+2)}{2(2k+1)}V_{m}\bar{V}_{n}\right.\\
\left.\left.-\frac{nm(n+2)\left(2m+1\right)}{2(2k+1)}V_{m}\bar{V}_{n}+\sum_{j=1}^{\infty}\left(\frac{C_{nj}}{j(j+1)}\right)\left(-\frac{mj(2m+j+6)}{2(2k+1)}V_{m}\bar{V}_{j}\right)\right]\right\} ,\label{eq:InPhaseMotionT_k}
\end{multline}
and
\begin{multline}
S_{k}=\delta\Re\left\{ \sum_{n=1}^{\infty}\sum_{m=1}^{\infty}\frac{k(k+2)(n+2)}{2(n+1)(2k+1)}\left(\bar{V}_{n}V_{m}\right)f_{knm}-\sum_{n=0}^{\infty}\sum_{m=0}^{\infty}\frac{k(k+2)m(m+2)}{2(2k+1)}\left(V_{m}\bar{V}_{n}\right)g_{knm}\right.\\
+\sum_{n=0}^{\infty}\sum_{m=1}^{\infty}a_{knm}\left[-\left(\frac{n^{2}}{4}+\frac{9n}{4}-\frac{5m^{2}}{4}-\frac{5m}{4}+\frac{1}{2}k(k-1)\right)\frac{m(m+2)}{2(2k+1)}V_{m}\bar{V}_{n}\right.\\
\left.\left.+\frac{nm(n+2)\left(2m+1\right)}{2(2k+1)}V_{m}\bar{V}_{n}+\sum_{j=1}^{\infty}\left(\frac{C_{nj}}{j(j+1)}\right)\left(\frac{mj(2m+j+6)}{2(2k+1)}V_{m}\bar{V}_{j}\right)\right]\right\} .\label{eq:InPhaseMotionS_k}
\end{multline}

\subsection{Discussion}

Our results show thus that a bubble, for any in-phase oscillation of its shape,  generates a streaming
of $O(\delta)$ or lower, and therefore  at least one order of magnitude weaker
than that of a  deformable no-slip surface (such as an 
elastic membrane) undergoing
the same sequence of  shape change. Thus the net flows generated by the angular velocities are of similar magnitude to those induced by the radial
velocities and they cancel at leading order. 

Longuet-Higgins observed that a bubble undergoing  translational oscillation
produced a force of $O(\delta)$, one order of magnitude less than
the out-of-phase translational and pulsating oscillations which is
$O(1)$ \cite{Higgins1998_Bubble}. Our calculations allow us to generalise this result to all shape changes, and thus suggests that  ensuring there are at least two modes of oscillation out of phase leads to stronger streaming flows.

For  bubbles forced by external fields, this  raises an interesting
question of whether a resonance mode of oscillation, which is solely
at one mode, would produce a weaker streaming flow than out-of-phase
forcing which excites multiple modes. From a practical standpoint, microbubbles are often fixed to a wall, which enforces that the center of the bubble has
to move and as such is naturally  excited at a second mode.

\section{Comparison with
 past work \label{part:ComparisonRileyHiggins}}

Our calculations have allowed us to  compute the streaming generated by any specified, fixed,
oscillating spherical object (and in particular we solved for a bubble). Past work has characterized the streaming flow for simple shape oscillations of bubbles and rigid spheres, to which we can compare our model in order to validate it. 

\subsection{Translating Bubble}

\begin{figure}[t]
\centering{}\subfloat[Streaming from current model.]{\protect\centering{}\protect
\includegraphics[scale=0.43]{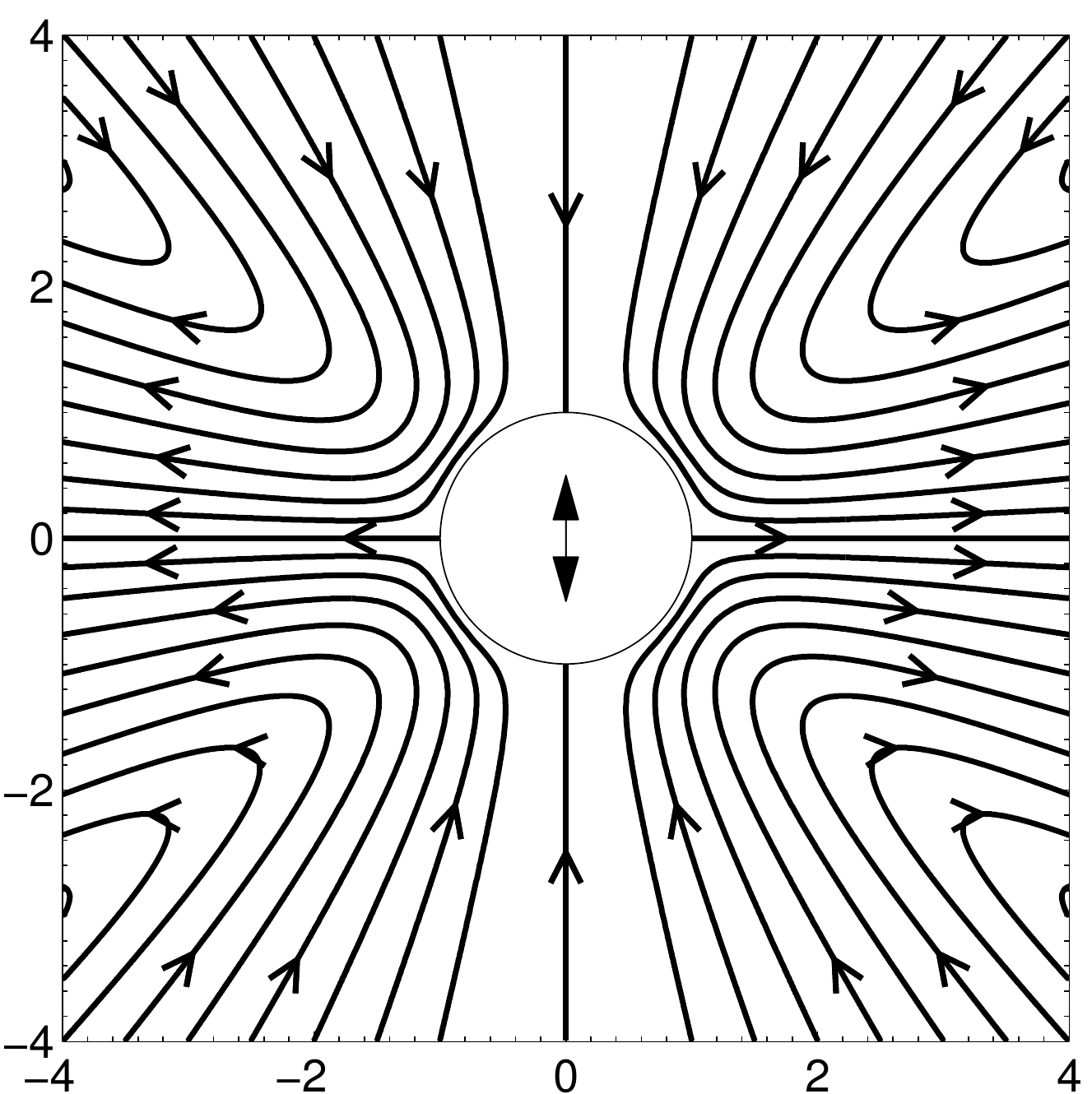}}
$\,\,\,\,\,\,\,\,\,\,\,\,\,\,\,\,$\subfloat[Longuet-Higgins's streaming \cite{Higgins1998_Bubble}.]{\protect\centering{}\protect\includegraphics[scale=0.55]{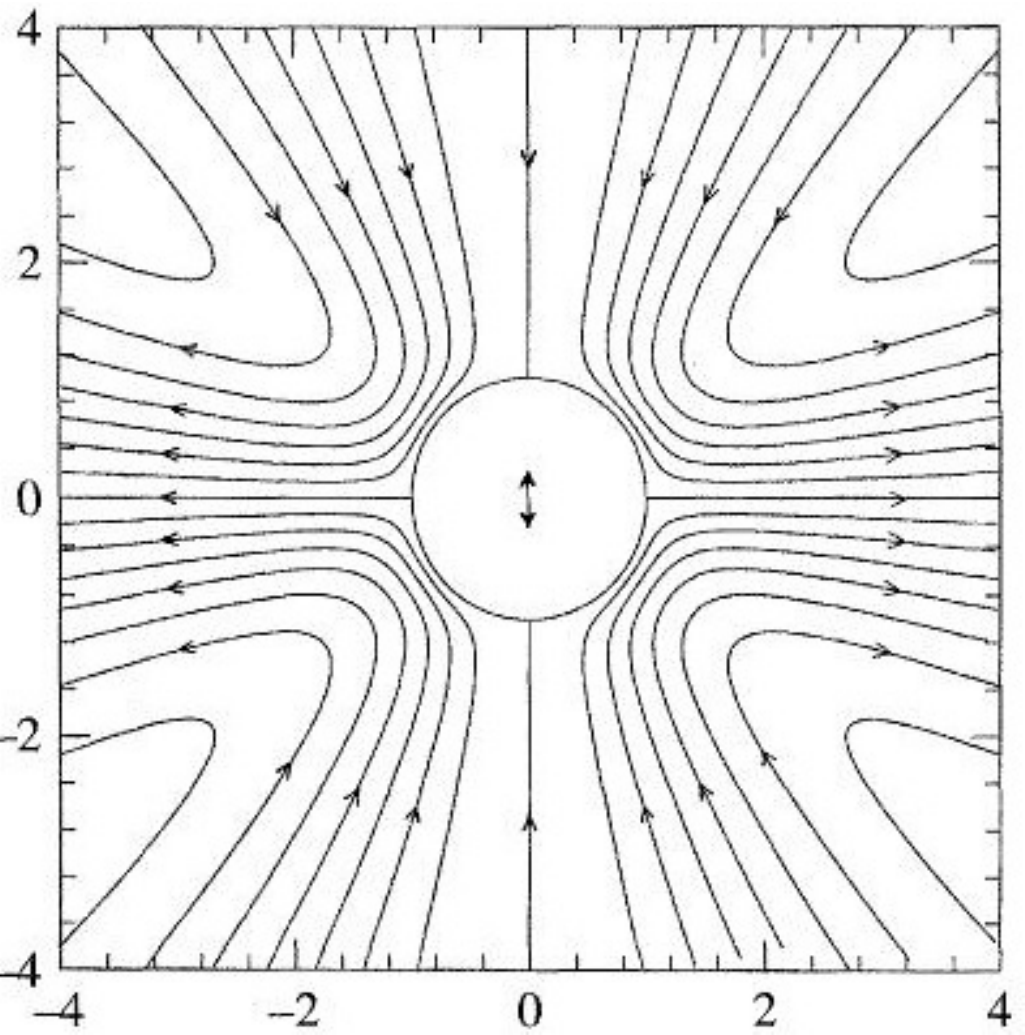}\protect}\protect\caption{\label{fig:LonguetHigginsInPhase}Streaming generated by the translational
oscillations of a bubble}
\end{figure}

In the case of a bubble undergoing translational oscillations, we have $V_{1}=1$ and  $V_{n}=0$ for $n\ne1$, The angular boundary conditions $W_{n}$
and $F_{n}$ are determined by the no stress boundary condition. This
case was studied by Longuet-Higgins \cite{Higgins1998_Bubble} and
the solution we obtain here is identical to his, namely \begin{equation}
 \langle \psi_{L} \rangle =\delta\frac{27}{20}\left(\frac{1}{r^{2}}-1\right)\int_{\mu}^{1}P_{2}(x)dx,\end{equation}
as further illustrated in Figure \ref{fig:LonguetHigginsInPhase}.

\subsection{Translating Sphere}
\begin{figure}[b]
\centering{}\subfloat[\label{fig:MyRiley} Streaming from current model.]{\protect\centering{}\protect
\includegraphics[scale=0.43]{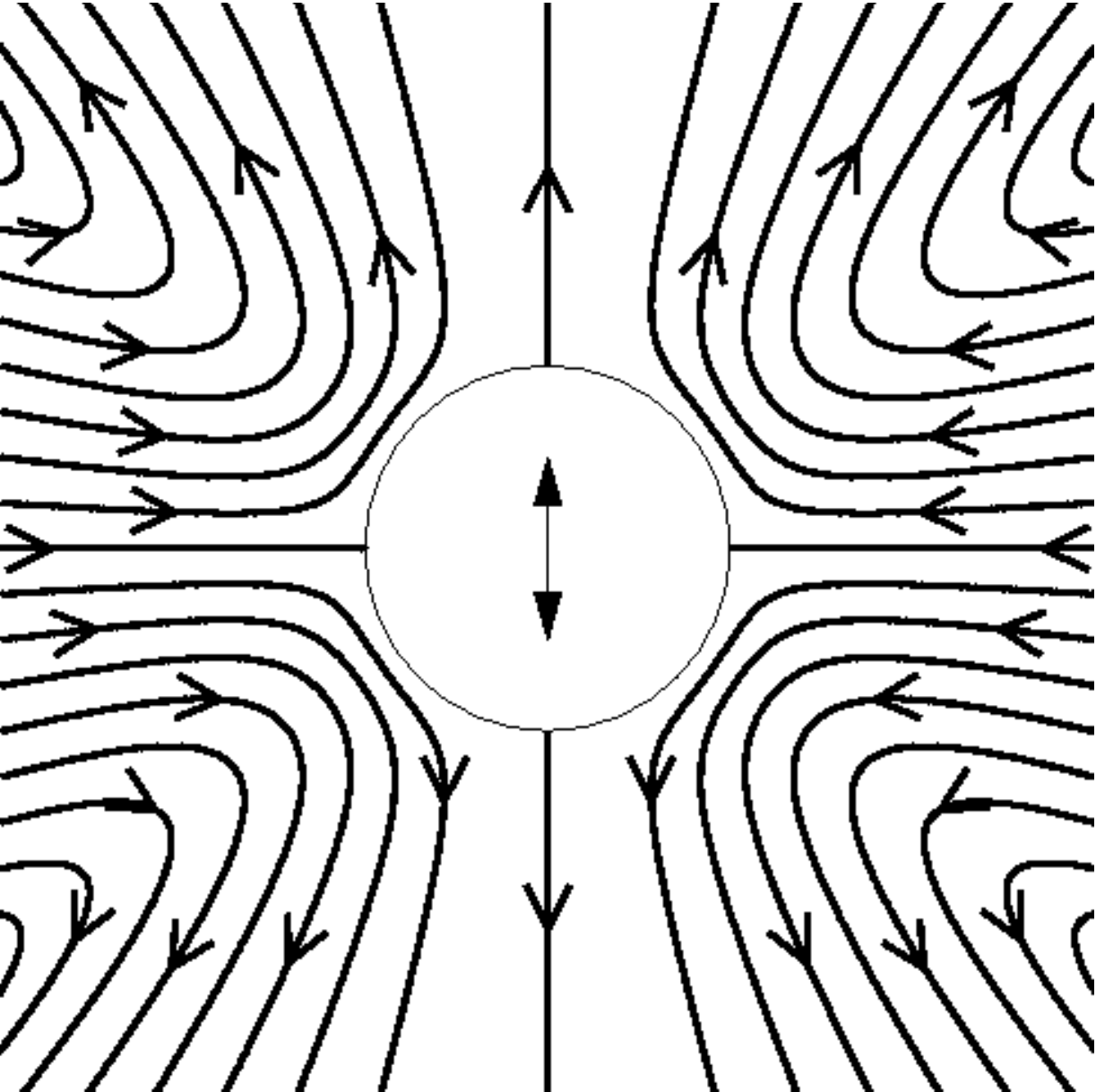}
\protect}
$\,\,\,\,\,\,\,\,\,\,\,\,\,\,\,\,\,\,\,$\subfloat[Riley's streaming \cite{Riley1966_SphereOscillating}.]{\protect\centering{}\protect\includegraphics[scale=0.4]{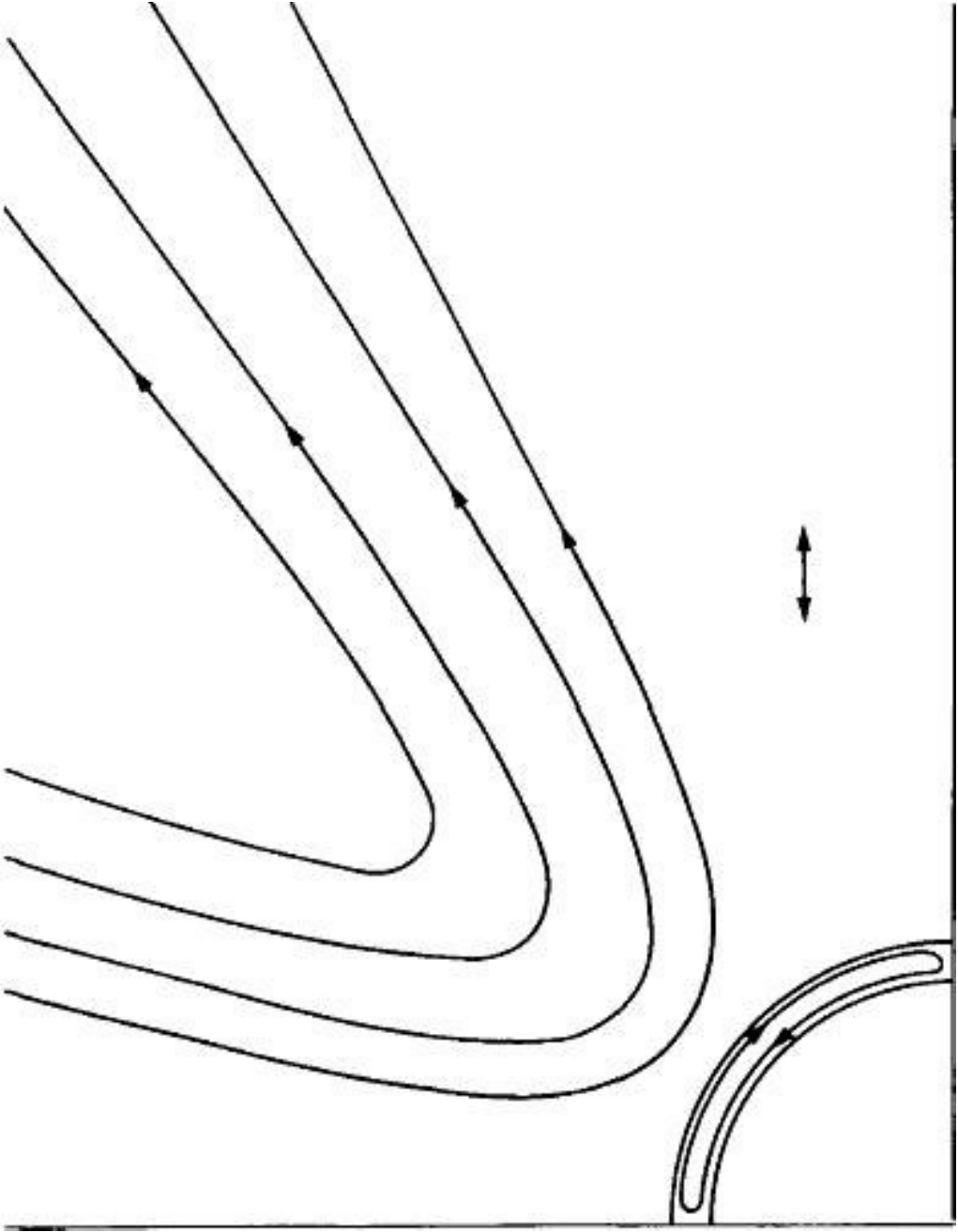}\protect}\protect\caption{\label{fig:Riley}Streaming generated by the translational oscillations
of a sphere }
\end{figure}

In  the case of a solid sphere undergoing translational oscillations we have 
$V_{1}=1$, $V_{n}=0$ for $n\ne1$, $W_{1}=2$, $W_{n}=0$ for $n\ne1$, and 
$F_{n}=0$. This case was studied by Riley \cite{Riley1966_SphereOscillating}
and the streaming we obtain is identical to his solution, namely 
\begin{equation}
\langle \psi_{2} \rangle =-\frac{45}{16}\left(\frac{1}{r^{2}}-1\right)\int_{\mu}^{1}P_{2}(x),
\end{equation}
as further illustrated in Figure \ref{fig:Riley}.

\subsection{Bubble Translating and Radially Oscillating}

Finally, in the case of a bubble undergoing radial and lateral oscillations
only $V_{0}$ and $V_{1}$ can be non-zero. Then the pairs $W_{n}$
and $F_{n}$ are determined by the no stress boundary condition. We obtain $W_{1}=-V_{1}$, $W_{n}=0$ for $n\ne1$, $F_{1}=4i\bar{V}_{0}V_{1}$ 
and $F_{n}=0$ for $n\ne1$. In this case there is a contribution
from the Stokes Drift with one non-zero component of $Y$, $Y_{101}={i\bar{V}_{0}V_{1}}/{2}$
leading to the final Lagrangian streaming as 
\begin{equation}
\langle \psi_{L} \rangle=\Re\left(i\bar{V}_{0}V_{1}\right)\left(-\frac{1}{4r}+\frac{r}{2}-\frac{1}{4r^{4}}\right)(1-\mu^{2}),
\end{equation}
which matches the result of Longuet-Higgins \cite{Higgins1998_Bubble} (note that in Ref.~\cite{Higgins1998_Bubble}, $\Re\left(i\bar{V}_{0}V_{1}\right)$
is written as $\sin(\phi)$ with $\phi$ denoting the phase difference between  modes 0 and 1).

\section{Illustration of steady streaming and far-field behaviour\label{part:StreamingExamples}}

With the calculations above, we can now illustrate the streaming patterns which can be obtained from surface oscillations. We consider a range of surface boundary conditions ($V_{n}$ and $W_{n}$) and assume for simplicity that the  surface streaming is   zero ($F_{n}=0$). The steady streaming flow splits naturally into two regions with different behaviours: the fluid motion close to the spherical body, which often contains recirculation regions, and the far-field behaviour which is dominated
by the slowest decaying term in the velocity.

\begin{figure}[t]
\subfloat[Oscillating with $V_{1}=1$ only. \label{fig:V_1_Only}]{\protect\includegraphics[width=0.3\textwidth]{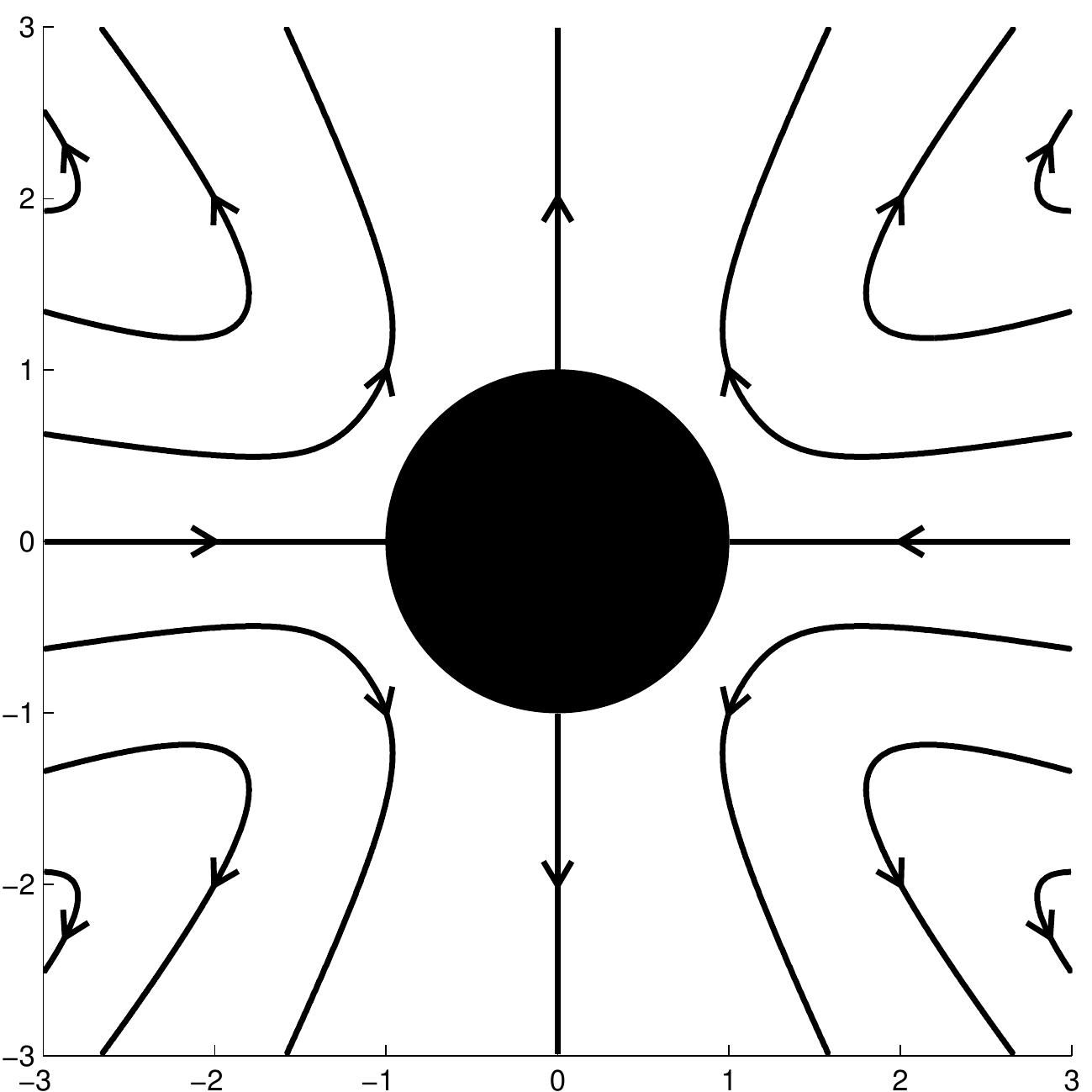}}
\, \,
\subfloat[Oscillating with $V_{2}=1$ only. \label{fig:V_2_Only}]{\protect\includegraphics[width=0.3\textwidth]{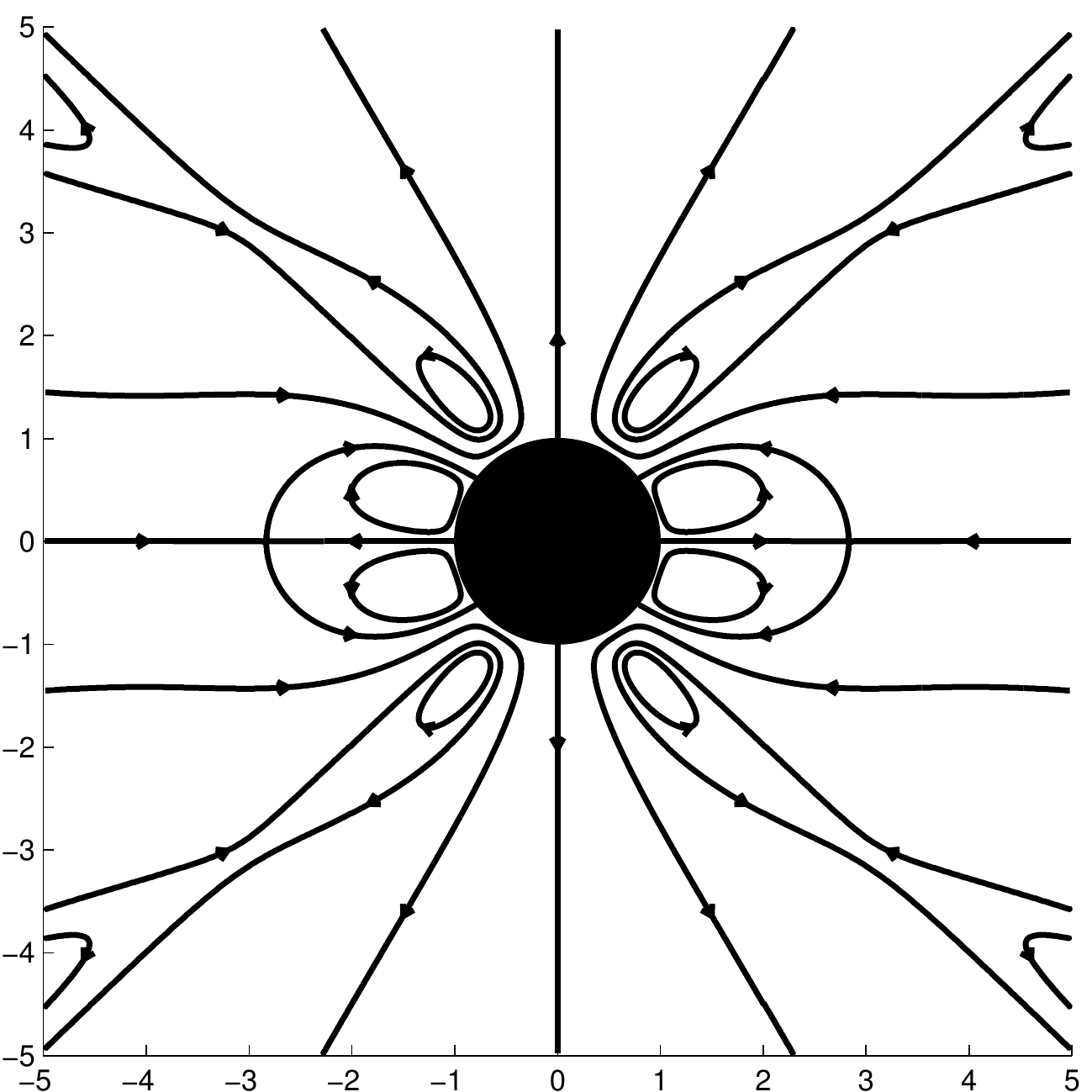}
}

\subfloat[Positive force: oscillating at two modes $V_{1}=\frac{1}{\sqrt{2}},$ $V_{2}=\frac{1}{\sqrt{2}},$ $W_{1}=\frac{-3}{\sqrt{2}}$
and $W_{2}=\frac{2}{\sqrt{2}}.$\label{fig:Force_V_1=000026V_2_OtherDirection}]{\protect\includegraphics[width=0.3\textwidth]{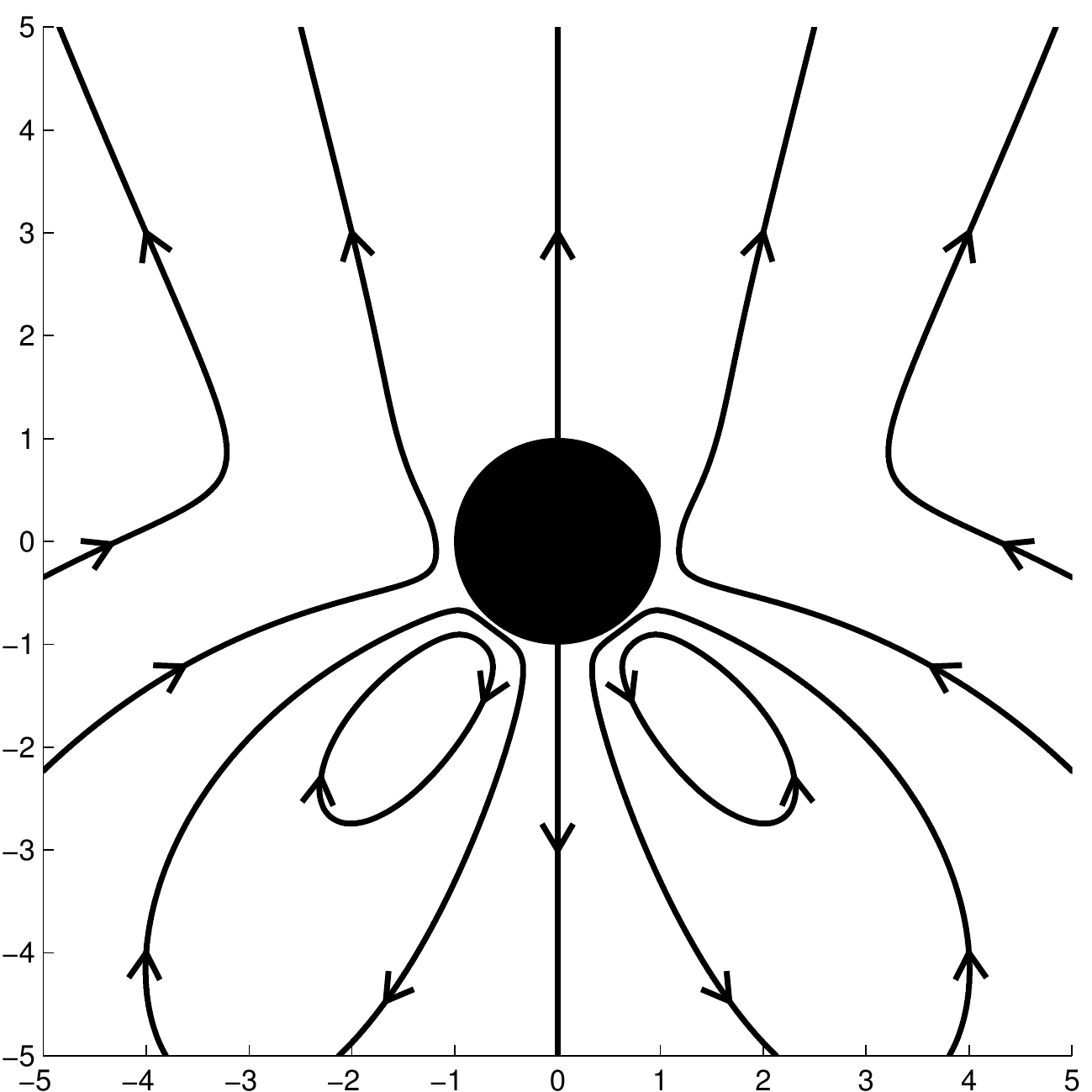}
}
\, \,
\subfloat[Negative force : oscillating at two modes $V_{1}=\frac{1}{\sqrt{2}}$ and $V_{2}=\frac{1}{\sqrt{2}}$.\label{fig:Force_V_1=000026V_2}]{\protect
\includegraphics[width=0.3\textwidth]{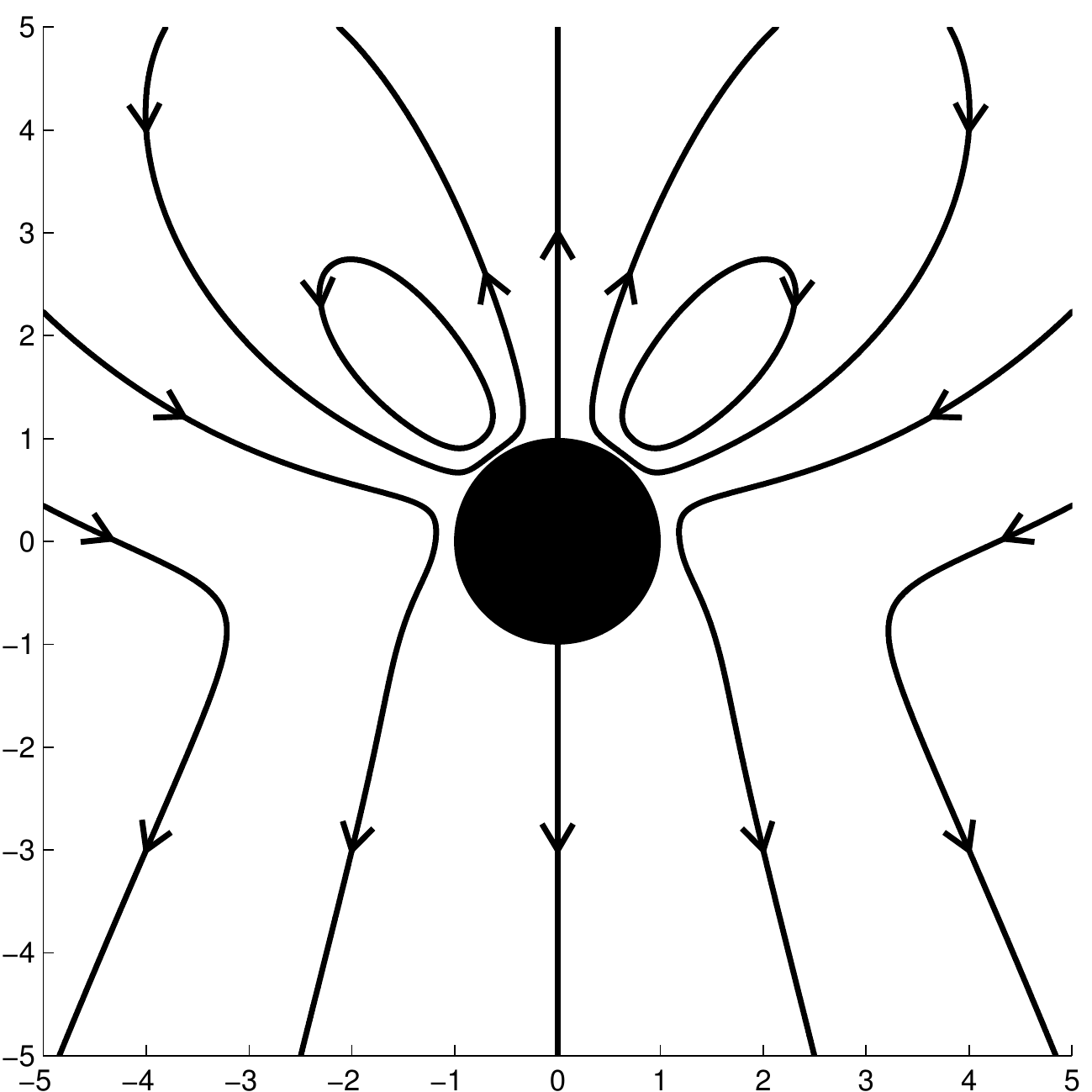}
}
\protect\caption{Patterns of steady streaming for the first few  surface  oscillation modes.}
\end{figure}

First we consider the streaming generated by a few simple surface
shape oscillations. In Figure \ref{fig:V_1_Only} we illustrate the streaming
when only the mode $V_{1}\neq 0$ is being forced. In all cases, the flow being axisymmetric, we only need to display streamlines in the plane of symmetry to illustrate the whole flow. Here $S_{1}=0$ so the slowest decaying term is $ {S_{2}}/{r^{2}}$ which produces the pattern of flow coming in in the equatorial plane of the spherical body and pushed away along the vertical axis (stresslet, see below). Furthermore, as there are few higher order terms this behaviour in fact dominates the flow throughout the domain. 

Differences between the far-field flow and the fluid motion close to the body
can be seen with higher modes. This is illustrated in Figure \ref{fig:V_2_Only}
which shows the streaming generated by forcing mode $V_{2}$ only. On the
edge of the figure the dynamics seen in figure \ref{fig:V_1_Only} is
apparent as the term ${S_{2}}/{r^{2}}$  is still dominant in the
far field. However close to the spherical body, we see circulations zones
which extend about one body diameter into the fluid. The number of
these circulations regions increase as higher modes are being forced. A similar pattern is  observed when only oscillating at one angular mode (i.e.~$W_{n}\neq0$ for one choice of $n$ ).

Let us now consider the  behaviour in the far field. Far from the sphere, the steady streaming is dominated by the slowest decaying
term. If $S_{1}\ne0$ then the slowest decaying flow is the Stokeslet
with the velocities decaying as ${S_{1}}/{r}$. This produces a non-zero force along the axis of rotational symmetry and this gives a clear movement of flow parallel
to this axis - either in the positive (Figure \ref{fig:Force_V_1=000026V_2_OtherDirection}) 
 or negative direction (Figure \ref{fig:Force_V_1=000026V_2}). 
If $S_{1}=0$ the body has no force acting on it, as seen
in Figures~\ref{fig:V_1_Only} and \ref{fig:V_2_Only} where the symmetry
of the system prevents a net force from being induced. Generally, a net force is created only if  two adjacent modes $\{V_{n},V_{n+1}\}$ or $\{W_{n},W_{n+1}\}$
are non-zero, as otherwise $a_{knm}$, $g_{knm}$, $f_{knm}$ are all
zero and the $a_{knm}c_{nj}$ combinations cancel.  

If the Stokeslet coefficient, $S_1$, is zero the far field behaviour is dominated
by a slower decaying term. In most cases it will be a stresslet with associated  velocities decaying as ${S_{2}}/{r^{2}}$. This is the flow seen  in Figure \ref{fig:V_1_Only}. If in turn these terms are also zero then the far-field behaviour is dominated by the $({T_{k}+S_{k+2}})/{r^{k+2}}$ term for the lowest value
of $k\geq 1$ which is non-zero.

Close to the spherical body, circulation regions will  form. If there is a Stokeslet,
this term tends to dominate the flow and even close to the body  circulation regions rarely appear. The one exception is close to the axis of symmetry where a pair of large circulations will
sometimes form, either just above or just below the sphere (see Figure
\ref{fig:Force_V_1=000026V_2} and \ref{fig:Force_V_1=000026V_2_OtherDirection}). Since the flow is axisymmetric, this corresponds to a recirculation torus. 

Intuitively, one would expect  that the shape of the volume physically displaced by
the spherical body (set by the modes $V_{n}$) would have a significant effect
on the features of the flow field. But as discussed  in
\S \ref{part:SPECIALCASE_bubble}, the angular motion on the surface  can produce a flow of similar magnitude, and in fact it can change the direction of the circulations and the streaming flow
(and hence the direction of the body force applied to the spherical
object) as demonstrated in Figures \ref{fig:Force_V_1=000026V_2} and
\ref{fig:Force_V_1=000026V_2_OtherDirection}.  Another example of
this, already seen in \S \ref{part:ComparisonRileyHiggins}, displayed
the streaming flow difference between a translating bubble (Figure
\ref{fig:LonguetHigginsInPhase}) and a translating sphere in (Figure
\ref{fig:Riley}).

\section{Application: force generation and propulsion\label{part:ForceAndVelocity}}

In our current setup, the sphere is held fixed, and the force exerted by the oscillations of the spherical body's surface on the fluid is computed. However, by  Newton's law, an equal and opposite force is being applied to the spherical body from the fluid. If the spherical body is not held in place then this would cause it to move. Mathematically, a net motion of the body is necessarily a second-order effect as all leading-order effects are oscillatory and produce no net motion or forces. Hence allowing the body to move will only slightly modify our mathematical approach. In this section we characterise  the force induced by a fixed body and then show
 how to carry the calculation in the case where the body is free to move.

\subsection{Force generation\label{sub:ForceGeneration}}

Due to the axisymmetry of the system a net force can only be exerted along
the axis of rotational symmetry, taken to be  ${\bf e}_{z}$ using traditional notation from spherical coordinates. We thus write $\boldsymbol{F}=F{\bf e}_{z}$.

The time averaged force on the spherical body is equal to the force
across the boundary of our spherical object at $r=R$, i.e.
\begin{equation}
\boldsymbol{F}=\left<\int\left(\boldsymbol{\sigma}\cdot\boldsymbol{n}\right)|_{r=R(\mu)}dS \right >.
\end{equation}
This force must match the force across the boundary ``$r=\infty$''
and thus 
\begin{equation}
\boldsymbol{F}=-\left < \int\left(\boldsymbol{\sigma}\cdot\boldsymbol{n}\right)|_{r=\infty}dS \right >.
\end{equation}
 Across $r=\infty$, if the time average is taken then there will
only be a contribution from the second order term as all first-order
terms are oscillatory. Also, the region $r=\infty$ is now
outside the boundary layer, where, due to the exponential decay of
the non-linear terms, the time-averaged \foreignlanguage{british}{behaviour}
is a Stokes flow. Outside the boundary layer, the slowest decaying velocity is the ${1}/{r}$ term. This velocity field at leading
order in $r$ is 
\begin{flalign}
u_{r} & =S_{1}\frac{\cos \theta}{r}+O(r^{-3}),\\
u_{\theta} & =-S_{1}\frac{\sin \theta}{2r}+O(r^{-3}).
\end{flalign}
This is the Stokeslet discussed above. Indeed
a non-dimensional Stokeslet due to a force $\boldsymbol{F}$ applied at the origin induces a flow $\boldsymbol{U}$ with  components
\begin{equation}
{U}_{j}=\frac{F_{i}}{8\pi}\left(\frac{\delta_{ij}}{r}+\frac{x_{i}x_{j}}{r^{3}}\right).
\end{equation}
With a force
 in the ${\bf e}_{z}$ direction this becomes gives 
\begin{eqnarray}
U_{r} & = & \frac{F}{4\pi r}\cos \theta,\\
U_{\theta} & = & -\frac{F}{8\pi r}\sin \theta.
\end{eqnarray}
Equating these two forms of the Stokeslet shows the non-dimensional
force exerted on the spherical body by the \foreignlanguage{british}{surrounding}
fluid is 
\begin{equation}
\boldsymbol{F}=-8\pi\epsilon^{2}\frac{S_{1}}{2}\boldsymbol{e}_{z}.
\end{equation}
This force is the result of a dominant pressure field, thus the dimensional
scaling for the pressure indicates the scaling for the force. The
Navier-Stokes equations indicate that the pressure scales with time
varying inertia so $p\sim\rho Ua\omega\sim\rho a^{2}\omega^{2}$ implying
that the dimensional force is 
\begin{equation}
\boldsymbol{F}=-4\pi\epsilon^{2}S_{1}\left(\rho a^{4}\omega^{2}\right)\boldsymbol{e}_{z}.\label{eq:Force}
\end{equation}

\subsection{Force-free swimming}

If the spherical body is no longer held in place but is free to move, it will translate with an $O(\epsilon^{2})$ velocity in the direction of this force. However, the constraint of force-free motion needs to be carefully enforced at both  $O(\epsilon)$ and $O(\epsilon^{2})$ and being free to
move will in general  also impact the first-order oscillatory motion.

\subsubsection{Force-free motion at $O(\epsilon)$ }

At $O(\epsilon)$ the motion of the spherical body is completely determined 
by the constants $V_{n}$ and $W_{n}$. Up to now these coefficients could be chosen  arbitrarily to represent any surface motion.  The extra constraint of force-free motion will now  restrict the allowed motion of the spherical body, therefore restricting choices of $V_{n}$ and $W_{n}$.

Mathematically, force-free motion is written as
\begin{equation}
\boldsymbol{F}=\int\left(\boldsymbol{\sigma}\cdot\boldsymbol{n}\right)|_{r=R(\mu,t)}dS=0.
\end{equation}
The normal vector to the surface of the spherical body is 
\begin{equation}
\boldsymbol{\hat{n}}=\boldsymbol{e}_{r}-\frac{\partial R}{\partial\theta}\boldsymbol{e}_{\theta}+O(\epsilon^{2}).\label{eq:NormalVector}
\end{equation}
Knowing that the direction of $\bf F$ is   in the ${\bf e}_{z}$ direction
by symmetry, this becomes 
\begin{equation}
\boldsymbol{F}=\left[\int\left(\sigma_{rr}\cos \theta-\sigma_{\theta r}\sin \theta\right)|_{r=1}dS+O(\epsilon^{2})\right]\boldsymbol{e}_{z}.
\end{equation}
This can then be non-dimensionalised and the integral expanded to give
\begin{equation}
\boldsymbol{F}=\left\{\int_{0}^{2\pi}\int_{0}^{\pi}\left[\left(-p+\delta^{2}\frac{\partial u_{r}}{\partial r}\right)\cos \theta-\frac{\delta^{2}}{2}\left(\frac{1}{r}\frac{\partial u_{r}}{\partial\theta}+\frac{\partial u_{\theta}}{\partial r}-\frac{u_{\theta}}{r}\right)\sin \theta \right]\sin \theta \,d\theta \,d\phi+O(\epsilon^2)\right\}\boldsymbol{e}_{z}.
\end{equation}
The first-order pressure can be calculated by substituting the first-order solution for $\psi_{1}$, (\ref{eq:FirstOrderSolution}), into the
Navier Stokes equation to obtain 
\begin{equation}
p=e^{it}\sum_{n=0}^{\infty}P_{n}(\mu)\left(\frac{iD_{n}}{(n+1)r^{n+1}}\right).
\end{equation}
 Then notice, using integration by parts and Legendre identities, 
that 
\begin{eqnarray}
\int_{0}^{\pi}P_{n}(\mu)\cos \theta \sin \theta d\theta&=&\frac{2}{3}\delta_{1n}\text{ for }n\ge0,\\
\int_{0}^{\pi}(1-\mu^{2})^{-\frac{1}{2}}{\left(\int_{\mu}^{1}P_{n}(x)dx\right)}\sin^{2}\theta \, d\theta&=&\frac{2}{3}\delta_{1n}\text{ for }n\ge1.
\end{eqnarray}
Therefore the force-free condition will only affect the $n=1$ mode.
Then as we have the scalings
\begin{equation}
\frac{\partial u_{r}}{\partial r}\sim O(1),\quad
\frac{\partial u_{r}}{\partial\theta}\sim O(1),\quad
\frac{\partial u_{\theta}}{\partial r}\sim O(\delta^{-1}), \quad
u_{\theta}\sim O(1),\quad p\sim O(1),
\end{equation}
at the two leading orders in $\delta$ only pressure
and the viscous stress $\sim {\partial u_{\theta}}/{\partial r}$ will contribute to the
force, leading to 
\begin{eqnarray}
\boldsymbol{F} & = & \frac{4\pi}{3}e^{it}\left[-\frac{iV_{1}}{2}+\frac{\delta(1+i)(W_{1}+V_{1})}{4}+O(\delta^{2})\right]{\bf e}_z.
\end{eqnarray}

For the spherical body to be force-free we thus need $V_{1}=W_{1}=0$. In other words, if the boundary conditions are such that $V_{1}\neq0$ and $W_{1}\neq0$
for a fixed sphere, then the force-free sphere will undergo additional oscillatory motion to compensate and lead to  $V_{1}=W_{1}=0$ overall. 
Physically the $V_{1}$ mode corresponds to the sphere undergoing
translational oscillations. So for the body to be force free the translational
oscillations are suppressed. At $O(\epsilon)$, since the behaviour
is linear, for most angular surface boundary conditions, $W_{1}$
will be directly dependent on the value of $V_{1}$ so a condition
restricting translational oscillations would be anticipated to effect
the angular motion at this mode too. 

\subsubsection{Force-free motion at $O(\epsilon^{2})$}

In order for the motion of the sphere at $O(\epsilon)$ to be force-free, we saw that two of the surface coefficients become zero. Beyond that, the  model has not fundamentally changed. We can thus use our mathematical framework to calculate  the velocity of translation at $O(\epsilon^{2})$ in terms of the force generated by the oscillating body which was force-free at first order.

We   use $\tilde{V}$ to denote the non-dimensional time-averaged  velocity
of the body at order $\epsilon^{2}$.  In order to use the same formulation as above, we
move into a frame of \foreignlanguage{british}{reference} where the
body undergoes no net motion at $O(\epsilon^{2})$ in the z direction. Mathematically,
this keeps all the boundary conditions the same as above except now
requires in the far field that
\begin{equation}
\psi\sim-\epsilon^{2}\tilde{V}\frac{r^{2}}{2}(1-\mu^{2}),\quad r\rightarrow\infty.\label{eq:MovingBodyCondition}
\end{equation}
As this   is an external boundary condition, the
form of the second-order inner solution  remains unchanged.
Therefore the main change is in the second-order outer solution. The
general form of this external streaming is still given by (\ref{eq:GeneralOuterFormulation})
but applying the new boundary condition (\ref{eq:MovingBodyCondition})
allows one more term than before to be non-zero so we have
\begin{multline}
\bar{\psi}_{2}=\tilde{R}_{0}\left(\int_{\mu}^{1}P_{0}(x)dx\right)+(-\tilde{V}r^{2}+\tilde{T}_{1}r^{-1}+\tilde{S}_{1}r)\left(\int_{\mu}^{1}P_{1}(x)dx\right)\\
+\sum_{n=2}^{\infty}\left(\tilde{T}_{n}r^{-n}+\tilde{S}_{n}r^{-(n-2)}\right)\left(\int_{\mu}^{1}P_{n}(x)dx\right).
\end{multline}
Then when comparing this outer solution to the inner solution (\ref{eq:L_k to T_k}),
(\ref{eq:M_k to T_k}) and (\ref{eq:R_0 to L_0}) all still hold for
$k\ne1$ but for $k=1$ we instead have
\begin{equation}
L_{1}=\tilde{T}_{1}+\tilde{S}_{1}-\tilde{V},
\end{equation}
\begin{equation}
\frac{M_{1}}{\delta}=\tilde{S}_{1}-\tilde{T}_{1}-2\tilde{V}.
\end{equation}

In order to determine the value of  $\tilde{V}$ we enforce that  the time-averaged
second-order solution be force-free so  we have 
\begin{equation}
\boldsymbol{F}=\left < \int\left(\boldsymbol{\sigma}\cdot\boldsymbol{n}\right)|_{r=R(\mu,t)}dS \right > ={\bf 0}.
\end{equation}
Contributions to this integral will come from linear terms involving
the internal second-order solution, $\psi_{2}$, as well as non-linear terms
involving the first order solution $\psi_{1}$.

As the form of the  first-order solution has not changed, the  contributions from those non-linear terms remain the same as when we restrict $\tilde{V}=0$ (i.e.~no second order translation).  However the second-order internal solution will give a slightly different contribution as the values of the constants of integration $N_{k}$ and $Q_{k}$ have changed. 

Looking at the second-order internal solution only the constants of
integration $L_{k}$, $M_{k}$, $N_{k}$ and $Q_{k}$ could be different. But changes will only occur for $k=1$ since for other values of $k$
the external solution is as before. Furthermore, the constants 
$L_{1}$ and $M_{1}$ are the same as before since their values were determined by the boundary conditions on the surface of the body which are the same. 
We use the asymptotic
matching in order to determine the values of $T_{1}$ and $S_{1}$ in terms of the
known values $L_{1}$ and $M_{1}$ and this is then used to calculate  the new values of $N_{1}$ and $Q_{1}$. 

In this new system we have
\begin{eqnarray}
L_{1}&=&\left(\tilde{T}_{1}+\frac{\tilde{V}}{2}\right)+\left(\tilde{S}_{1}-\frac{3\tilde{V}}{2}\right),
\\
\frac{M_{1}}{\delta}&=&\left(\tilde{S}_{1}-\frac{3\tilde{V}}{2}\right)-\left(\tilde{T}_{1}+\frac{\tilde{V}}{2},\right),
\\
\frac{N_{1}}{\delta^{2}}&=&\tilde{T}_{1}-\tilde{V}=\left(\tilde{T}_{1}+\frac{\tilde{V}}{2}\right)-\frac{3\tilde{V}}{2},
\\
\frac{Q_{1}}{\delta^{3}}&=&-\tilde{T}_{1}=-\left(\tilde{T}_{1}+\frac{\tilde{V}}{2}\right)+\frac{\tilde{V}}{2}\cdot
\end{eqnarray}
Compared to $S_{1}$ and $T_{1}$ in the first-order force-free case,
$T_{1}=\tilde{T}_{1}+{\tilde{V}}/{2}$ and $S_{1}=\tilde{S}_{1}-{3\tilde{V}}/{2}$. 
Therefore the drag force on the sphere will be the same as before,
$-4\pi\mu S_{1}\boldsymbol{e}_{z}$ plus an extra contribution from
the $-{3\tilde{V}}/{2}$ extra term in $N_{1}$ and the ${\tilde{V}}/{2}$
term in $Q_{1}$. 

Knowing $\hat{\boldsymbol{n}}$ from (\ref{eq:NormalVector}) and
that the direction of $F$ is still in the $\boldsymbol{e}_{z}$ direction means we have
\begin{equation}
\boldsymbol{F}=\left< \int\left[\left(\sigma_{rr}-\epsilon\frac{\partial R}{\partial\theta}\sigma_{\theta r}\right)\cos\theta-\left(\sigma_{\theta r}-\epsilon\frac{\partial R}{\partial\theta}\sigma_{\theta\theta}\right)\sin\theta \right]|_{r=R(\mu,t)}dS\right> \boldsymbol{e}_{z}.
\end{equation}
We see that the extra contributions to $F$ can only come from the linear terms
evaluated at $r=1$ (so $\eta=0$) i.e. 
\begin{equation}
\sigma_{rr}\cos \theta-\sigma_{\theta r}\sin \theta =\left[\left(-p+\delta^{2}\frac{\partial u_{r}}{\partial r}\right)\cos \theta -\frac{\delta^{2}}{2}\left(\frac{1}{r}\frac{\partial u_{r}}{\partial\theta}+\frac{\partial u_{\theta}}{\partial r}-\frac{u_{\theta}}{r}\right)\sin \theta\right].
\end{equation}

In $\sigma_{\theta r}$ there will only be an extra $N_{1}$ contribution
arising from the ${\partial u_{\theta}}/{\partial r}$ term. Any extra
contribution from $\sigma_{rr}$ will come from the pressure term.
The non-dimensional second-order equation for the pressure is 
\begin{equation}
\epsilon\left< \boldsymbol{u}\cdot\nabla\boldsymbol{u} \right > =-\nabla\left<\boldsymbol{p} \right>+\frac{\delta^{2}}{2}\mu\nabla^{2}\left<\boldsymbol{u} \right>,
\end{equation}
where the pressure scales as $p\sim\rho aU\omega$. Then $\left<\boldsymbol{u}\cdot\nabla \boldsymbol{u} \right>$ gives a contribution in terms
of $\psi_{1}$ only. There is however an extra contribution coming from $\nabla^{2}\left<\boldsymbol{u} \right> $
. By looking at the $\theta$ component of this equation, and noticing
that we are evaluating at $r=1$ in the integral, we see we are only interested
in terms with no dependence of $\eta.$ Then it can be found that
the $N_{k}$ and $Q_{k}$ contribution to $p$ is given by
\begin{equation}
p=\frac{3}{2}\delta^{2}\left(\frac{Q_{1}}{\delta^{3}}\right)P_{1}(\mu).
\end{equation}
The extra contribution to $F$ coming from the ${\partial u_{\theta}}/{\partial r}$ term 
in $\sigma_{\theta r}$ and from the pressure is 
\begin{equation}
\delta^{2}\int_{0}^{2\pi}\int_{0}^{\pi} \left[ \left(-\frac{3\tilde{V}}{4}\right)\sin^{3}\theta-\left(\frac{3\tilde{V}}{4}\right)\cos^{2} \theta \sin \theta \right] d\theta d\phi=-3\pi\delta^{2}V.
\end{equation}
Therefore the total non-dimensional force on the spherical body is 
\begin{equation}
{\bf F} = \epsilon^{2}\left(-4\pi S_{1}-3\pi\delta^{2}\tilde{V}\right)\boldsymbol{e}_{z}.
\end{equation}
For the body to be force free its dimensional velocity must therefore be 
\begin{equation}\label{122}
\tilde{V}=-\frac{4}{3\delta^{2}}(a\omega)S_{1},
\end{equation}
which we can use to calculate a numerical value for $\tilde{V}$.

The relationship between the force on the body and its velocity is given by substituting this value of $S_{1}$ into (\ref{eq:Force}) giving 
\begin{equation}
\boldsymbol{F}=3\pi\epsilon^{2}\delta^{2}\tilde{V}(\rho a^{3}\omega)\boldsymbol{e}_{z}.
\end{equation}
Then substituting the scaling for $\delta^{2}$ given by (\ref{eq:delta_boundarylayersize})
finally leads to  
\begin{equation}
\boldsymbol{F}=6\pi\epsilon^{2}\tilde{V}\left(a\rho\nu\right)\boldsymbol{e}_{z}.
\end{equation}
We recognise  the standard result for a solid sphere translating at speed
$\epsilon^2\tilde{V}$ in a Stokes flow. Since outside the boundary layer the
flow is a Stokes flow, such a similarity was in fact  expected (corrections for non-sphericity are expected at higher orders in $\epsilon$).

\section{Conclusion}

In this paper we have mathematically derived the steady streaming flow generated by arbitrary axisymmetric shape oscillations of a spherical body. The final solution, and thus the main result of this paper, is quantified in 
Eqs.~\eqref{eq:FinalSolutionEquation}-\eqref{eq:Tt}.

 Our model, which agrees with classical results, shows that a  net force is generated in the far field only when two adjacent surface modes are excited.   If the body is free to move, this force will cause the body to move with a net velocity, which we derived, given by a balance between that streaming force and the Stokes drag (Eq.~\ref{122}). 
 
 Having kept the boundary forcing arbitrary makes our model  applicable to  a wide range of microorganisms and  microfluidic devices. Recent experimental work  has used bubbles embedded in free-moving hollow bodies to generate propulsion \cite{Fend_Micropropulsion_2015}. Our formalism will be directly applicable to this new class of synthetic swimmers and could help improve future designs \cite{BertinSwimmer_2015}.

\section*{Acknowledgements}
This work was funded in part by the EPSRC (TS) and the European Union through a Marie Curie CIG Grant (EL).

\appendix

\section{Out-of-phase streaming around a bubble\label{chap: AppendixA}}

In order to apply the general steady streaming model (\ref{eq:FinalSolutionEquation})
specifically to a bubble the  boundary condition of tangential
stress on the boundary of the spherical body,
\begin{equation}
\boldsymbol{n}\cdot\boldsymbol{\sigma}\cdot\boldsymbol{t}=0\quad \text{at}\quad r=R,\,\,\mbox{\ensuremath{\theta}=\ensuremath{\Theta}},
\end{equation}
needs to be applied. This will determine the angular motion on the
surface of the bubble $W_{n}$ and $F_{n}$ in terms of a prescribed
radial motion $V_{n}$.

\subsection{Boundary condition}

We denote the unit tangent vector to the body's surface in the plane through the axis of axisymmetry $\boldsymbol{t}$ and the  normal vector $\boldsymbol{n}$. 
  Both can be calculated in terms of  ${\bf e}_{z}$ and 
${\bf e}_{\theta}$ measured from the center of the rest position
of the body giving
\begin{eqnarray}
\boldsymbol{t}&=&{\bf e}_{\theta}+\frac{\partial R}{\partial\theta}{\bf e}_{r}+O(\epsilon^{3}),\\
\boldsymbol{n}&=&{\bf e}_{r}-\frac{\partial R}{\partial\theta}{\bf e}_{\theta}+O(\epsilon^{3}).
\end{eqnarray}
Using these equations the no stress condition
can be expanded in terms of $\epsilon$ giving the first two terms
as
\begin{equation}
\sigma_{\theta r}+\frac{\partial R}{\partial\theta_{0}}\left(\sigma_{rr}-\sigma_{\theta\theta}\right)+O(\epsilon^{3})=0\quad \text{at}\quad r=R,\,\,\mbox{\ensuremath{\theta}=\ensuremath{\Theta}},
\end{equation}
becoming
\begin{equation}
\sigma_{\theta r}+(R-1)\frac{\partial\sigma_{\theta r}}{\partial r}+(\Theta-\theta_{0})\frac{\partial\sigma_{\theta r}}{\partial\theta}+\left(\sigma_{rr}-\sigma_{\theta\theta}\right)\frac{\partial R}{\partial\theta}+O(\epsilon^{3})=0
\quad \text{at}\quad r=1,\,\,\theta=\theta_{0}
.\label{eq:NoTangentialStressEquation}
\end{equation}

\subsection{Leading-order solution}

At $O(\epsilon)$, Eq.~\ref{eq:NoTangentialStressEquation} reduces
to the non-dimensional equation
\begin{equation}
\left(\frac{1}{r}\frac{\partial u_{r}}{\partial\theta}+\frac{\partial u_{\theta}}{\partial r}-\frac{u_{\theta}}{r}\right)=0\quad \text{at}\quad r=1,\,\,\theta=\theta_{0}.
\end{equation}
By substituting in the known first-order solution $\psi_{1}$, (\ref{eq:FirstOrderSolution}), 
this equation can be used to determine $W_{n}$  
giving 
\begin{equation}
W_{n}=-nV_{n}+\delta\left[\frac{2n(n+2)}{(1+i)}V_{n}\right]+\delta^{2}\left[in(n+2)^{2}V_{n}\right]+O(\delta^{3})\label{eq:No_Stress_Wm}.
\end{equation}

\subsection{Second-order solution}

At $O(\epsilon^{2})$, after
Taylor expansion of $\sigma_{\theta r}$ 
~\eqref{eq:NoTangentialStressEquation}  reduces to the non-dimensional
equation 
\begin{equation}
\sigma_{\theta r}^{(\epsilon^{2})}=-(R-1)\frac{\partial\sigma_{\theta r}^{(\epsilon)}}{\partial R}-(\Theta-\theta_{0})\frac{\partial\sigma_{\theta r}^{(\epsilon)}}{\partial\theta}-\frac{\partial R}{\partial\theta_{0}}\left(\sigma_{rr}^{(\epsilon)}-\sigma_{\theta\theta}^{(\epsilon)}\right)
\quad \text{at}\quad r=R,\,\,\mbox{\ensuremath{\theta}=\ensuremath{\Theta}},\label{eq:SecondOrderNoStressGeneralEquation}
\end{equation}
where superscripts indicate the order at which each term  is to be 
taken. Upon substitution of $R$ , $\Theta$ and $\sigma$,  (\ref{eq:SecondOrderNoStressGeneralEquation}) 
becomes 
\begin{multline}
\left((1-\mu^{2})\frac{\partial^{2}\psi_{2}}{\partial\mu^{2}}+2\frac{\partial\psi_{2}}{\partial r}-\frac{\partial^{2}\psi_{2}}{\partial r^{2}}\right)=\\
\Re\left[\frac{i}{\omega}\sum_{n=0}^{\infty}V_{n}P_{n}(\mu)e^{it}\right]\Re\left[\left((1-\mu^{2})\frac{\partial^{3}\psi_{1}}{\partial\mu^{2}\partial r}+3\frac{\partial^{2}\psi_{1}}{\partial r^{2}}-\frac{\partial^{3}\psi_{1}}{\partial r^{3}}-3(1-\mu^{2})\frac{\partial^{2}\psi_{1}}{\partial\mu^{2}}-4\frac{\partial\psi_{1}}{\partial r}\right)\right]\\
+\Re\left[\frac{\epsilon i}{\omega}\sum_{n=1}^{\infty}W_{n}\left(\int_{\mu}^{1}P_{n}(x)dx\right)e^{it}\right]\Re\left[\frac{\mu}{(1-\mu^{2})}\left((1-\mu^{2})\frac{\partial^{2}\psi_{1}}{\partial\mu^{2}}+2\frac{\partial\psi_{1}}{\partial r}-\frac{\partial^{2}\psi_{1}}{\partial r^{2}}\right)\right.\\
\left.+\left(\frac{\partial\left((1-\mu^{2})\frac{\partial^{2}\psi_{1}}{\partial\mu^{2}}\right)}{\partial\mu}+2\frac{\partial^{2}\psi_{1}}{\partial r\partial\mu}-\frac{\partial^{3}\psi_{1}}{\partial r^{2}\partial\mu}\right)\right]-\Re\left[\frac{\epsilon i}{\omega}\sum_{n=0}^{\infty}n(n+1)V_{n}\left(\int_{\mu}^{1}P_{n}(x)dx\right)e^{it}\right]\\
\times\left[2\left(3\frac{\partial\psi_{1}}{\partial\mu}-2\frac{\partial\psi_{1}}{\partial r\partial\mu}-\frac{\mu}{(1-\mu^{2})}\frac{\partial\psi_{1}}{\partial r}\right)\right]\\
\quad \text{at}\quad r=1,\,\,\theta=\theta_{0}.\label{eq:FullStressEquation}
\end{multline}
The  forcing on the 
right-hand side of (\ref{eq:FullStressEquation})  can be calculated explicitly from the derivatives
of $\psi_{1}$, (\ref{eq:FirstOrderSolution}), simplifying the equation
to 
\begin{multline}
\left \langle (1-\mu^{2})\frac{\partial^{2}\psi_{2}}{\partial\mu^{2}}+2\frac{\partial\psi_{2}}{\partial r}-\frac{\partial^{2}\psi_{2}}{\partial r^{2}} \right \rangle =\frac{i}{2}\sum_{k=0}^{\infty}\sum_{n=0}^{\infty}\sum_{m=1}^{\infty}a_{knm}\left\{ -m(m^{2}+2)V_{m}\bar{V}_{n}\phantom{\frac{c_{nj}}{(j+1)}}\right.\\
+(m+2)^{2}W_{m}\bar{V}_{n}-4m(m+1)\bar{W}_{n}V_{m}+\bar{V}_{n}(W_{m}+mV_{m})\left[\frac{2i}{\delta^{2}}+\frac{(m+3)(1+i)}{\delta}+\right.\\
\left.\frac{m(2m+3)}{2}\right]-W_{m}(\bar{W}_{n}+n\bar{V}_{n})\frac{(1-i)}{\delta}+W_{m}\bar{V}_{n}n(n+2)-(n+2)\bar{W}_{n}W_{m}\\
+\sum_{j=1}^{\infty}\left(\frac{C_{nj}}{j(j+1)}\right)\left[\frac{1}{\delta}\frac{W_{m}\left(\bar{W}_{j}+j\bar{V}_{j}\right)(1-i)}{j(j+1)}+\frac{(j+2)W_{m}\left(\bar{W}_{j}-j\bar{V}_{j}\right)}{j(j+1)}\right.\\
\left.\left.\phantom{\frac{c_{nj}}{(j+1)}}+2m(m+1)V_{m}\bar{W}_{j}\right]\right\} \int_{\mu}^{1}P_{k}(x)dx\label{eq:RHS}.
\end{multline}

The left hand side of (\ref{eq:FullStressEquation}) can be evaluated
using $\psi_{2}$ from (\ref{eq:InnerSolution}) and the value of the three
constants of integration, $L_{k}$, $M_{k}$ and $N_{k}$, are needed.
$L_{k}$ and $M_{k}$ were calculated in (\ref{eq:Lt-1}) and (\ref{eq:Mt}).
From the asymptotic matching $L_{k}$ and $M_{k}$ are known in terms
of $T_{k}$ and $S_{k}$, (\ref{eq:L_k to T_k}) and (\ref{eq:M_k to T_k}).
Matching at one higher order determines $N_{k}$ in terms of $T_{k}$
and $S_{k}$, then using (\ref{eq:L_k to T_k}) and (\ref{eq:M_k to T_k}), 
$N_{k}$ can be written in terms of $L_{k}$ and $M_{k}$ giving 
\begin{equation}
N_{k}=\left[\frac{(2-k)k}{2}L_{k}+\frac{(1-2k)}{2}\left(\frac{M_{k}}{\delta}\right)\right]\delta^{2}.\label{eq:N_k}
\end{equation}

Equation (\ref{eq:FullStressEquation}) leads to an equality for $\sum_{k=1}^{\infty}F_{k}$$\left(\int_{\mu}^{1}P_{k}(x)dx\right)$
at leading order, $O(1)$. As this must hold for all $\mu\in[-1,1]$,   the coefficients of $\left(\int_{\mu}^{1}P_{k}(x)dx\right)$ must
equate for each $k$, and thus this  will give an equation for every
$F_{k}$ separately. Upon substitution of (\ref{eq:No_Stress_Wm}) the equation for $F_{k}$
reduces to
\begin{multline}
\frac{F_{k}^{(\delta)}}{\delta}=-\sum_{n=0}^{\infty}\sum_{m=1}^{\infty}a_{knm}\left[m(m+4+n)\frac{i}{2}\bar{V}_{n}V_{m}+\sum_{j=1}^{\infty}\left(\frac{C_{nj}}{(j+1)}\right)\left(\frac{i}{2}m\bar{V}_{m}V_{j}\right)\right]\\
-\frac{3k}{(2k+1)}V_{0}\bar{V}_{k}i-\sum_{n=0}^{\infty}\sum_{m=1}^{\infty}\frac{9kg_{knm}}{2(2k+1)}V_{m}\bar{V}_{n}i+\sum_{n=1}^{\infty}\sum_{m=1}^{\infty}\frac{3kf_{knm}}{2(2k+1)}\frac{(n+2)i\bar{V}_{n}V_{m}}{(n+1)(m+1)}\\
+\sum_{n=0}^{\infty}\sum_{m=1}^{\infty}\frac{a_{knm}}{(2k+1)}\left[m\left(n^{2}-4nm-4n+m^{2}-m-3\right)i\bar{V}_{n}V_{m}\phantom{\frac{c_{nj}}{(j+1)}}\right.\\
\left.+\sum_{j=1}^{\infty}\left(\frac{C_{nj}}{(j+1)}\right)m(3m+5)iV_{m}\bar{V}_{j}\right].\label{eq:NoShearEquation}
\end{multline}
Then substituting $F_{k}^{(\delta)}$ into (\ref{eq:St}) and (\ref{eq:Tt}) finally
gives 
\begin{multline}
 T_{k}=\Re\left\{ \frac{(1-k^{2})}{(2k+1)}iV_{0}\bar{V}_{k}+\sum_{n=1}^{\infty}\sum_{m=1}^{\infty}\frac{\left(k^{2}-k-1\right)(n+2)}{2(2k+1)(n+1)(m+1)}f_{knm}i\bar{V}_{n}V_{m}\right.\\
+\sum_{n=0}^{\infty}\sum_{m=1}^{\infty}\frac{3(1-k^{2})}{2(2k+1)}iV_{m}\bar{V}_{n}g_{knm}+\sum_{n=0}^{\infty}\sum_{m=1}^{\infty}\frac{a_{knm}}{2(2k+1)}\left[m\left(n^{2}-4nm-4n\right.\phantom{\frac{c_{nj}}{(}}\right.\\
\left.\left.\left.+m^{2}-m-3\right)i\bar{V}_{n}V_{m}+\sum_{j=1}^{\infty}\left(\frac{C_{nj}}{j+1}\right)m(3m+5)iV_{m}\bar{V}_{j}\right]\right\} +O(\delta)\label{eq:T_k_bubble}
\end{multline}
and
\begin{multline}
S_{k}=\Re\left\{ \frac{k(k+2)}{(2k+1)}iV_{0}\bar{V}_{k}-\sum_{n=1}^{\infty}\sum_{m=1}^{\infty}\frac{\left(2k+4\right)k(n+2)}{4\left(2k+1\right)(n+1)(m+1)}f_{knm}i\bar{V}_{n}V_{m}\right.\\
+\sum_{n=0}^{\infty}\sum_{m=1}^{\infty}\frac{3k(k+2)}{2(2k+1)}iV_{m}\bar{V}_{n}g_{knm}-\sum_{n=0}^{\infty}\sum_{m=1}^{\infty}\frac{a_{knm}}{2(2k+1)}\left[m\left(n^{2}-4nm-4n\right.\phantom{\frac{c_{nj}}{(}}\right.\\
\left.\left.\left.+m^{2}-m-3\right)i\bar{V}_{n}V_{m}+\sum_{j=1}^{\infty}\left(\frac{C_{nj}}{j+1}\right)m(3m+5)iV_{m}\bar{V}_{j}\right]\right\} +O(\delta)\label{eq:S_k_bubble}
\end{multline}
which with (\ref{eq:Y_knm}) give the value of all the constants in
the streaming solution (\ref{eq:FinalSolutionEquation}).

\section{In-phase streaming around a bubble
\label{sec:Appendix2_InPhaseBubble}}

Notice, from the solution of the stedy streaming around a bubble
given in Appendix \ref{chap: AppendixA}, that all terms in (\ref{eq:S_k_bubble}),
(\ref{eq:T_k_bubble}) and (\ref{eq:Y_knm}) are multiplies of $iV_{m}\bar{V}_{k}$
for some integers $m,k\ge0$. As such Appendix \ref{chap: AppendixA}
only gives the solution for out-of-phase motion of the bubble, as
otherwise that solution is identically zero. Therefore for these cases, the steady streaming needs to be calculated to the next order, namely $O(\delta)$. 

\subsection{Inner-solution at third order in $\delta$}

In order to  find the steady streaming at $O(\delta)$, the asymptotic matching
must be carried out at one higher order. As such the inner second-order solution must be calculated to an extra order in $\delta$.
Therefore more terms will be required in the $\delta$ expansions,
so we first return to the second-order, inner governing equation
\begin{equation}
\frac{\delta^{2}}{2} \langle D^{4} \psi_{2} \rangle =\frac{1}{r^{2}} \left \langle \frac{\partial(\psi_{1},D^{2}\psi_{1})}{\partial(r,\mu)}+2L\psi_{1}D\psi_{1} \right \rangle.\label{eq:secondordergoverningequationproper-1}
\end{equation}
When expanding the left hand side in $\delta$, the $O(1)$ terms
in $\delta$ now contribute to the streaming as well as the $O(\delta^{-2})$
term so we have
\begin{equation}
\frac{\delta^{2}}{2} \langle D^{4} \psi_{2} \rangle =\frac{1}{2\delta^{2}}\frac{\partial^{4} \langle \psi_{2} \rangle }{\partial\eta^{4}}+(1-\mu^{2})\frac{\partial^{4} \langle \psi_{2} \rangle }{\partial\eta^{2}\partial\mu^{2}}+O(\delta).\label{eq:D^2_SecondExpansion}
\end{equation}
From (\ref{eq:InnerSolution}), the $O(1)$ (leading-order) solution
of $\langle \psi_{2} \rangle $ is known. The second term in (\ref{eq:D^2_SecondExpansion})
will only make an $O(1)$ or lower contribution from this term so
the governing equation for the first three orders of $ \langle \psi_{2} \rangle $ can
be simplified to
\begin{multline}
\frac{\partial^{4} \langle \psi_{2} \rangle}{\partial\eta^{4}}=\frac{2\delta^{2}}{r^{2}} \left \langle \frac{\partial(\psi_{1},D^{2}\psi_{1})}{\partial(r,\mu)}+2L\psi_{1}D\psi_{1}\right \rangle \\
+\delta^{2}\sum_{k=1}^{\infty}\sum_{n=0}^{\infty}\sum_{m=1}^{\infty}a_{knm}\left[2k(k+1)(W_{m}+mV_{m})\bar{V}_{n}e^{-(1+i)\eta}\right]\left(\int_{\mu}^{1}P_{k}(x)dx\right)+O(\delta^{3}).\label{eq:LongerInnerEquation}
\end{multline}
Expanding the term $\left \langle {\partial(\psi_{1},D^{2}\psi_{1})}/{\partial(r,\mu)}+2L\psi_{1}D\psi_{1}\right \rangle$ to
$O(\delta^{2})$ and substituting into (\ref{eq:LongerInnerEquation})
then gives an equation for $\left \langle {\partial^{4}  \psi_{2} }/ {\partial\eta^{4}} \right \rangle $
which can be integrated twice to find that the $O(\delta^{2})$ (only)
term in $ \left \langle {\partial^{2} \psi_{2}} /{\partial\eta^{2}} \right \rangle$
is 
\begin{multline}
\left \langle \frac{\partial^{2} \psi_{2}}{\partial\eta^{2}}\right \rangle ^{(\delta^{2})}=\sum_{k=1}^{\infty}\left(\int_{\mu}^{1}P_{k}(x)dx\right)\sum_{n=0}^{\infty}\sum_{m=1}^{\infty}a_{knm}\left\{ -m(\bar{W}_{n}+n\bar{V}_{n})V_{m}i\left(\frac{n}{2}-m-3\right)\right.\\
+(\bar{W}_{n}+n\bar{V}_{n})(W_{m}+mV_{m})\left(\frac{i}{2}\left(1-m+n\right)-(2+\frac{n}{2})\right)\\
+\left(W_{m}+mV_{m}\right)\bar{V}_{n}i\left(-\frac{5n^{2}}{4}-\frac{21n}{4}+nm+m+\frac{m^{2}}{4}-3-t(t+1)\right)\\
+\sum_{j=1}^{\infty}\left(\frac{C_{nj}}{j(j+1)}\right)\left[(\bar{W}_{j}+j\bar{V}_{j})(W_{m}+mV_{m})\left((1+i)-\frac{1}{2}\left(m+ij\right)+j\right)\right.\\
\left.\left.+\bar{V}_{j}(W_{m}+mV_{m})ij(2j+6-m)\right]\right\} .\label{eq:ThirdOrderDerivitive}
\end{multline}
The $O(\delta^{2})$ contribution to $\langle \psi_{2} \rangle$ can be calculated
by integrating twice more but in order to satisfy the no tangential stress
boundary condition only $\left \langle {\partial^{2}\bar{\psi}_{2}}/{\partial\eta^{2}}\right \rangle $
is required at $O(\delta^{2})$. 

Notice that every term in $\langle \psi_{2} \rangle $ up to and including $O(1)$
is a multiple of $W_{n}+nV_{n}$ for some $n$ so every term will
drop an order when the first order stress condition $W_{n}=-nV_{n}+O(\delta)$
is applied. Similarly, higher-order terms in $\langle \psi_{2} \rangle$ will also
be multiples of $W_{n}+nV_{n}$ since in $\left \langle {\partial(\psi_{1},D^{2}\psi_{1})}/{\partial(r,\mu)}+2L\psi_{1}D\psi_{1}\right \rangle $
every term is a multiple of $B_{n}\propto(W_{n}+nV_{n})$ and in $ \langle D^{4} \psi_{2} \rangle $
extra terms in its delta expansion will be in terms of lower orders
of $\langle \psi_{2} \rangle $ which are also proportional to $W_{n}+nV_{n}$. Therefore even when calculating the streaming to $O(\delta)$ for
a bubble, only the first three terms up to $O(1)$ are needed in the
equation for the inner streaming.

\subsection{Constants of integration}

The second order inner solution $\langle \psi_{2} \rangle $ is of the form 
\begin{equation}
\langle \psi_{2} \rangle =\sum_{k=1}^{\infty}\left(\int_{\mu}^{1}P_{k}(x)dx\right)\left[f_{k}(\eta)+g_{k}(\eta)\delta+h_{k}(\eta)\delta^{2}+L_{k}+M_{k}\eta+N_{k}\eta^{2}+O(\delta^{3})\right],
\end{equation}
 where $f$ and $g$ are known from (\ref{eq:InnerSolution}) and
$h$ could be found by integrating (\ref{eq:ThirdOrderDerivitive}).
The no-tangential stress boundary condition then gives 
\begin{multline}
\left \langle (1-\mu^{2})\frac{\partial^{2}\bar{\psi}_{2}}{\partial\mu^{2}}+2\frac{\partial\bar{\psi}_{2}}{\partial r}-\frac{\partial^{2}\bar{\psi}_{2}}{\partial r^{2}}\right \rangle =\sum_{k=1}^{\infty}\left\{ \frac{1}{\delta^{2}}\left(-\frac{\partial^{2}f_{k}}{\partial\eta^{2}}\right)+\frac{1}{\delta}\left(2\frac{\partial f_{k}}{\partial\eta}-\frac{\partial^{2}g_{k}}{\partial\eta^{2}}\right)\right.\\
\left.+\left[-k(k+1)f_{k}(\eta)+2\frac{\partial g_{k}}{\partial\eta}-\frac{\partial^{2}h_{k}}{\partial\eta^{2}}+2\frac{M_{k}}{\delta}-2\frac{N_{k}}{\delta^{2}}-k(k+1)L_{k}\right]\right\} \left(\int_{\mu}^{1}P_{k}(x)dx\right).\label{eq:NoTangentialStressOneSide}
\end{multline}
The $O(\delta^{-2})$ terms will cancel with the $O(\delta^{-2})$
quantity in (\ref{eq:RHS}). The $O(\delta^{-1})$ terms do not cancel
exactly but when taking $W_{n}=-nV_{n}+O(\delta)$ (the first order
bubble condition) they do. Therefore the $O(1)$ terms will give the
leading-order behaviour for which the value of the constants $L_{k}$, $M_{k}$
and $N_{k}$ are needed. The constants $L_{k}$ and $M_{k}$ were calculated in (\ref{eq:Lt-1}) and (\ref{eq:Mt})
and $N_{k}$ can be written in terms of $L_{k}$ and $M_{k}$, (\ref{eq:N_k}).

Equation (\ref{eq:NoTangentialStressOneSide}) can then be equated with (\ref{eq:RHS})
to give the algebraic condition for no tangential stress. This will
give a condition on $M_{k}$ and $L_{k}$ but $L_{k}$ is uniquely
determined by the boundary condition: the radial velocity of the bubble
equals the radial velocity of the fluid adjacent to the bubble. $M_{k}$
was also determined but is a function of the unknown $F_{k}$ which
this no tangential stress condition will determine. However
$F_{k}$ uniquely determines $M_{k}$ so this equation can be considered
as just determining $M_{k}$ giving
\begin{multline}
-(2k+1)\left(\frac{M_{k}}{\delta}\right)=(-3k)L_{k}+\sum_{n=0}^{\infty}\sum_{m=1}^{\infty}a_{knm}\left(\frac{1}{\delta}\left\{ (W_{m}+mV_{m})(\bar{W}_{n}+n\bar{V}_{n})\frac{(i-1)}{2}\right.\right.\\
+(W_{m}+mV_{m})\frac{\bar{V}_{n}}{2}(1-i)\left(2n+3\right)+\sum_{j=1}^{\infty}\left(\frac{C_{nj}}{j(j+1)}\right)\left[(\bar{W}_{j}+j\bar{V}_{j})(W_{m}+mV_{m})\frac{(1+i)}{2}\right.\\
\left.\left.-(W_{m}+mV_{m})\bar{V}_{j}j(1-i)-\frac{W_{m}\left(\bar{W}_{j}+j\bar{V}_{j}\right)(1+i)}{2j(j+1)}\right]\right\} +\left\{ m(\bar{W}_{n}+n\bar{V}_{n})V_{m}i\left(\frac{n}{2}-m-2\right)\right.\\
-(\bar{W}_{n}+n\bar{V}_{n})(W_{m}+mV_{m})\left[\frac{i}{2}\left(n-m\right)+\left(\frac{n}{2}-\frac{1}{2}\right)\right]-\left(W_{m}+mV_{m}\right)\bar{V}_{n}i\left[-\frac{5n^{2}}{4}-\frac{9n}{4}+nm\right.\\
\left.+\frac{m^{2}}{4}+1-\frac{3}{2}k(k+1)+\frac{(2m^{2}+3m)}{4}\right]-\frac{i}{2}W_{m}\bar{V}_{n}n(n+2)+\frac{i}{2}(n+2)\bar{W}_{n}W_{m} \\
%+\frac{i}{2}m(m^{2}+2)V_{m}\bar{V}_{n}\\
%-\frac{i}{2}W_{m}\bar{V}_{n}n(n+2)+\frac{i}{2}(n+2)\bar{W}_{n}W_{m}
+\frac{i}{2}m(m^{2}+2)V_{m}\bar{V}_{n}-\frac{i}{2}\left[(m+2)^{2}W_{m}\bar{V}_{n}-4m(m+1)\bar{W}_{n}V_{m}\right]\\
+\sum_{j=1}^{\infty}\left(\frac{C_{nj}}{j(j+1)}\right)\left[-(\bar{W}_{j}+j\bar{V}_{j})(W_{m}+mV_{m})\left(\frac{(1+i)}{2}-\frac{1}{2}\left(m+ij\right)+j\right)\right.\\
\left.\left.\left.-\bar{V}_{j}(W_{m}+mV_{m})ij(2j+4-m)-im(m+1)V_{m}\bar{W}_{j}-\frac{i}{2}\left(\frac{(j+2)W_{m}\left(\bar{W}_{j}-j\bar{V}_{j}\right)}{j(j+1)}\right)\right]\right\} \right)+O(\delta).\label{eq:nostresscondition}
\end{multline}

When applying the first-order stress boundary condition (\ref{eq:No_Stress_Wm})
all terms drop by one order in $\delta$. Then assuming the $O(1)$
terms cancel (which is required for the result in Appendix \ref{chap: AppendixA}  not to
give the solution) this gives $M_{k}$ at $O(\delta)$. Then $L_{k}$ can
be calculated at $O(\delta)$ by applying (\ref{eq:RadialCondition_SecondOrderProper})
at $O(\delta)$ to $\langle \psi_{2} \rangle $ (\ref{eq:InnerSolution}). This gives
\begin{multline}
L_{k}=\delta\left\{ \sum_{n=0}^{\infty}\sum_{m=1}^{\infty}a_{knm}\frac{m(m+2)}{2}\left[(1+i)V_{m}\bar{V}_{n}\right]-\sum_{n=0}^{\infty}\sum_{m=0}^{\infty}\frac{m(m+2)}{2}\left[(1+i)V_{m}\bar{V}_{n}\right]g_{knm}\right.\\
\left.+\sum_{n=1}^{\infty}\sum_{m=1}^{\infty}\frac{(n+2)(m+2)f_{knm}}{2(n+1)(m+1)}\left[(1+i)\bar{V}_{n}V_{m}\right]-\sum_{m=1}^{\infty}\sum_{n=1}^{\infty}\frac{(n+2)}{2(n+1)(m+1)}f_{knm}\left[(1-i)\bar{V}_{n}V_{m}\right]\right\} 
,\end{multline}
and the simplified $M_{k}$ expression 
\begin{multline}
\left(\frac{M_{k}}{\delta}\right)=\frac{\left(3k\right)}{(2k+1)}L_{k}-\delta\sum_{n=0}^{\infty}\sum_{m=1}^{\infty}\frac{a_{knm}}{(2k+1)}\left\{ -nm(n+2)\left(2m+1\right)(1-i)V_{m}\bar{V}_{n}\right.\\
+m(m+2)\left[\frac{n^{2}}{4}+\frac{9n}{4}-\frac{5m^{2}}{4}+\frac{3}{2}t(t+1)-\frac{5m}{4}\right](1+i)V_{m}\bar{V}_{n}\\
\left.+\sum_{j=1}^{\infty}\left(\frac{C_{nj}}{j(j+1)}\right)\left[mj(m+1)(j+2)(1-i)V_{m}\bar{V}_{j}-(j+4)mj(m+2)(1+i)V_{m}\bar{V}_{j}\right]\right\}.
\end{multline}

\subsection{Outer streaming constants}

Using the matching conditions (\ref{eq:T_K to M_k value}) and
(\ref{eq:S_k to M_k value}) we finally obtain
\begin{multline}
T_{k}=\delta\Re\left(\sum_{n=1}^{\infty}\sum_{m=1}^{\infty}\frac{(1-k^{2})(n+2)}{2(2k+1)(n+1)(m+1)}f_{knm}\left[(m+2)(1+i)\bar{V_{n}}V_{m}-(1-i)\bar{V}_{n}V_{m}\right]\right.\\
-\sum_{n=0}^{\infty}\sum_{m=0}^{\infty}\frac{(1-k^{2})m(m+2)}{2(2k+1)}\left[(1+i)V_{m}\bar{V}_{n}\right]g_{knm}+\sum_{n=0}^{\infty}\sum_{m=1}^{\infty}a_{knm}\left\{ m(m+2)\left(\frac{n^{2}}{4}+\frac{9n}{4}\right.\right.\\
\left.-\frac{5m^{2}}{4}-\frac{5m}{4}+1+\frac{1}{2}k^{2}+\frac{3}{2}k\right)\frac{(1+i)}{2(2k+1)}V_{m}\bar{V}_{n}-nm(n+2)\left(2m+1\right)\frac{(1-i)}{2(2k+1)}V_{m}\bar{V}_{n}\\
\left.\left.+\sum_{j=1}^{\infty}\left(\frac{C_{nj}}{(j+1)}\right)\left[m(m+1)(j+2)\frac{(1-i)}{2(2k+1)}V_{m}\bar{V}_{j}-(j+4)m(m+2)\frac{(1+i)}{2(2k+1)}V_{m}\bar{V}_{j}\right]\right\} \right),
\end{multline}
and
\begin{multline}
S_{k}=\delta\Re\left(\sum_{n=1}^{\infty}\sum_{m=1}^{\infty}\frac{k(k+2)(n+2)}{2(n+1)(m+1)(2k+1)}f_{knm}\left[(m+2)(1+i)\bar{V}_{n}V_{m}-(1-i)\bar{V}_{n}V_{m}\right]\right.\\
-\sum_{n=0}^{\infty}\sum_{m=0}^{\infty}\frac{k(k+2)m(m+2)}{2(2k+1)}\left[(1+i)V_{m}\bar{V}_{n}\right]g_{knm}+\sum_{n=0}^{\infty}\sum_{m=1}^{\infty}a_{knm}\left\{ -m(m+2)\left[\frac{n^{2}}{4}+\frac{9n}{4}\right.\right.\\
\left.-\frac{5m^{2}}{4}-\frac{5m}{4}+\frac{1}{2}k(k-1)\right]\frac{(1+i)}{2(2k+1)}V_{m}\bar{V}_{n}+nm(n+2)\left(2m+1\right)\frac{(1-i)}{2(2k+1)}V_{m}\bar{V}_{n}\\
\left.\left.+\sum_{j=1}^{\infty}\left(\frac{C_{nj}}{(j+1)}\right)\left[-m(m+1)(j+2)\frac{(1-i)}{2(2k+1)}V_{m}\bar{V}_{j}+(j+4)m(m+2)\frac{(1+i)}{2(2k+1)}V_{m}\bar{V}_{j}\right]\right\} \right),
\end{multline}
which, with $Y_{knm}=0$, give the value of all the constants
in the streaming solution (\ref{eq:OuterSolution}). 

\bibliographystyle{unsrt}
\bibliography{References}

\end{document}